\documentclass[a4paper,12pt]{article}

\usepackage{latexsym}
\usepackage{amsmath,amssymb,amsxtra,amscd,amsthm}
\usepackage[mathscr]{eucal}

\DeclareFontFamily{U}{rsfs}{\skewchar\font127 }
\DeclareFontShape{U}{rsfs}{m}{n}{
   <5> rsfs5
   <6> rsfs6
   <7> rsfs7
   <8> rsfs8
   <9> rsfs9
   <10> rsfs10
   <10.95> rsfs11
   <12> rsfs12
   <14.4> rsfs14
   <17.28> rsfs17
   <20.74> rsfs20
   <24.88> rsfs25
   <29.86-> rsfs30}{}
\DeclareMathAlphabet\scr{U}{rsfs}{m}{n}
\def\cO{\mathcal{O}}

\def\mC{\mathbb{C}}

\def\mP{\mathbb{P}}

\def\mR{\mathbb{R}}

\def\mT{\mathbb{T}}

\def\mZ{\mathbb{Z}}

\def\tL{\mathrm{L}}

\def\tS{\mathrm{S}}

\def\tl{\mathrm{l}}

\def\ts{\mathrm{s}}

\newfont{\HUGE}{cmssbx10 scaled 4000}

\DeclareMathOperator{\chern}{ch}

\DeclareMathOperator{\todd}{td}
\DeclareMathOperator{\p}{p}

\def\bb1{\textup{\small{1}} \kern-3.8pt \textup{1}}
\def\SL2Z{\tS\tL(2,\mZ)}

\numberwithin{equation}{section}

\providecommand{\href}[2]{#2}

\newtheoremstyle{plain} % name
  {5pt}%      Space above
  {10pt}%      Space below
  {\rmfamily} %         Body font
  {}%         Indent amount (empty = no indent, \parindent = para indent)
  {\scshape} % Thm head font
  {}%        Punctuation after thm head
  {\newline} %     Space after thm head: " " = normal interword space;
  {}%         Thm head spec (can be left empty, meaning `normal')
\swapnumbers
\theoremstyle{break}
\renewcommand{\p}{\partial}
\def\gammat{\widetilde{\gamma}}
\def\betat{\widetilde{\beta}}
\def\a{\alpha}
\def\b{\beta}
\def\g{\gamma}
\def\l{\lambda}

\newcommand\cdr[1]{\Omega^{\mathrm{ch}}_{#1}}
\newcommand\css[1]{\cO^{\mathrm{ch}}_{#1}}
\def\Ell{\mathrm{Ell}}

\begin{document}
\begin{titlepage}
\begin{flushright}
DISTA-2007\\
LPTENS 07/06\\
hep-th/0702044
\end{flushright}
\vskip 1.5cm
\begin{center}
{\LARGE \bf 
Partition Functions, Localization, 
\\\vskip .2cm
and the Chiral de Rham complex
} 
\vfill
{\large Pietro Antonio Grassi$^{1,2}$, Giuseppe Policastro$^3$, and Emanuel Scheidegger$^2$ } \\
\vfill {
$^1$ Centro Studi e Ricerche E. Fermi,
Compendio Viminale, I-00184, Roma, Italy,\\
\vskip .2cm
$^{2}$ Dipartimento di Scienze e Tecnologie Avanzate, Universit\`a del Piemonte Orientale\\
via Bellini 25/g, 15100 Alessandria, Italy, and INFN - Sezione di
Torino, Italy\\
\vskip .2cm
$^3$
Laboratoire de Physique Th\'eorique de l'\'Ecole Normale Sup\'erieure\\
24, Rue Lhomond - 75231 Paris Cedex 05, France
}
\end{center}
\vfill
\begin{abstract}
We propose a localization formula for the chiral de Rham complex
generalizing the well-known localization procedure in topological
theories. Our formula takes into account the contribution due to the
massive modes. The key to achieve this is to view the non-linear
$\beta\gamma$ system as a gauge theory. For abelian gauge groups we are in
the realm of toric geometry. Including the $bc$ system, the formula
reproduces the known results for the elliptic genus of toric varieties. We
compute the partition function of several models.
\end{abstract}
\vfill

\date{February 2007}
\end{titlepage}

\tableofcontents

\section{Introduction}
\label{sec:intro}

Some of the recent developments in string theory have been based on 
the formulation of topological string theory. This is indeed an important sector 
of string theory and it captures some of the geometrical information of Calabi-Yau compactifications to four dimensions in a physical string theory. Topological string theory is defined by the twisted version of the $N=2$ superconformal algebra generated by the worldsheet fields and related to the Calabi-Yau geometrical data; 
this excludes from the Fock space all massive modes of the physical string spectrum. Of course, the complete spectrum of the theory on a Calabi-Yau manifold would be a very important achievement since it would allow a more precise description of low-energy physics. However, since currently there are no techniques available that permit a complete analysis of this spectrum, the best we can do is to consider a chiral model on the same Calabi-Yau manifold. This construction is known as half-twisted (0,2) sigma model~\cite{Witten:1991zz}, \cite{Witten:2005px}. The underlying structure is again a twisted superconformal algebra, but here the definition of the physical states does not exclude the infinite tower of massive states for the left-movers. Therefore, it gives new hints about the spectrum of string theory on Calabi-Yau manifolds and it stands between the topological and the physical string theory. 

The information of the spectrum is encoded in the one-loop twisted partition function  
\begin{equation*}
Z_X(t, y| q) = {\mathrm Tr}_{{\cal F}_X}\left( (-y)^{F_0} t^{J_0} q^{L_0}\right)\,,
\end{equation*}
where $X$ is the target space manifold, $F_0$ is the zero mode of a fermionic current, $J_0$ is the zero mode of a bosonic $U(1)$ current and $L_0$ is the 
zero mode of the Virasoro generator. ${\cal F}_X$ is the Fock space.
The first term $Z^0_X(t, y| q)$ of the $q$ expansion of $Z_X(t, y| q)$ is the contribution of the zero modes and coincides with the topological string approximation (except for the worldsheet instantons since they are absent in the large volume limit~\cite{Kapustin:2005pt}, \cite{Witten:2005px}). Notice that in the definition of $Z_X(t, y| q)$, the fermionic current $F$ enters. This is related to the presence of fermionic degrees of freedom on the worldsheet. In this case $Z_X(t, y| q)$ is called the elliptic genus of the manifold~\cite{Witten:1986bf} (see below for a precise definition). Moreover, the elliptic genus is a topological invariant of the manifold and therefore it has a geometrical meaning~\cite{Hirzebruch:1992ab}. The contribution of the zero modes encodes 
some of the geometrical informations, but the elliptic genus carries 
the information about the conformal invariance, the reparametrization invariance in terms of its modular invariance. 

The half-twisted (0,2) model has a beautiful realization in terms of non-linear $\beta\gamma$ systems~\cite{Witten:2005px}. By half-twisting the (2,2) model one obtains a supersymmetric non--linear sigma model written 
in terms of commuting fields $\beta,\gamma$ (and anticommuting fields $b,c$). To be more precise, in the case of (0,2) models, the BRST cohomology is equivalent to the space of vertex operators of the $\beta\gamma$ system on each open set on the manifold (in the mathematical literature~\cite{Malikov:1998dw}, \cite{Malikov:1999ab}, \cite{Malikov:1999ac}, \cite{Gorbounov:1999ab}, \cite{Gorbounov:2000ab},\cite{Gorbounov:2000ac} this is known as the chiral structure sheaf). In the case of $(2,2)$ models the BRST cohomology is equivalent to a $\beta\gamma bc$ system (see~\cite{Witten:2005px}, \cite{Kapustin:2005pt}) and the model is denoted in \cite{Malikov:1998dw} as chiral de Rham algebra or chiral de Rham complex (which are sheaves, too). The fields $\beta,\gamma,b,c$ are holomorphic free fields on the worldsheet whose target space is a single patch of the manifold $X$. The physical information is encoded in the gluing procedure needed to extend the local vertex operators to globally-defined ones. In the case of supersymmetric models, there is an identification between the physical spectrum of (0,2) half-twisted model and the non-linear sigma model, however the analysis can be performed for a non-linear $\beta\gamma$ system even without the supersymmetric completion. In the supersymmetric case, all anomalies vanish, while in the bosonic case they vanish if $p_{1}(X)=0$ and $c_{1}(X) c_{1}(\Sigma)=0$ where $\Sigma$ is the Riemann surface, $c_{1}(X)$ and $p_{1}(X)$ are the first Chern class and the first Pontrjagin class of $X$, respectively.  

In the very interesting paper~\cite{Borisov:1998dw} the structure of the chiral de Rham complex is investigated completely from the BRST point of view. The authors realize that a very convenient setting to construct specific examples is toric geometry. In this context, Borisov explicitly shows how to go from the free theory to the theory after the gluing over the complete set of patches. In further work~\cite{Borisov:1999ab},~\cite{Borisov:2000fg} the elliptic genus for the chiral de Rham theory has been determined for compact toric varieties, non-compact toric varieties which are local Calabi--Yau manifolds, and finally compact Calabi--Yau manifolds given as hypersurfaces in toric varieties.
The elliptic genera are constructed by studying the toric data and gluing the local information to a globally defined modular invariant quantity. Less is known about the partition function in the bosonic case of the chiral structure sheaf.  In addition, the formulas and the prescription given in the above papers appear to be difficult to grasp. 

There is another important motivation to study the non-linear $\beta\gamma$ systems: the ghost system of the pure spinor string theory \cite{Berkovits:2000fe}, \cite{Grassi:2001ug}, \cite{Grassi:2002tz} is one instance of such a system and it is characterized by 11 independent worldsheet fields $\l^{\a}$ (and their conjugates $w_{\a}$) living on the quadric $\l^{\a} \g_{\a\b}^{m}\l^{\b}=0$ (or equivalently living on the coset $\mathrm{SO}(10) \times \mC^{*}/ \mathrm{SU}(5)$, $\l^{\a}$ are called pure spinors). The complete pure spinor string theory, in the flat background, is composed of three sectors 1) 10 free bosonic standard worldsheet fields, 2) 16 free fermionic $(1,0)$ fields and 3) the non-linear sector of the pure spinors. The computation of the spectrum boils down to compute the spectrum of the pure spinor sector since the rest is very simple~\cite{Berkovits:2005hy,Grassi:2005jz,Grassi:2006wh}. 
 
Here we present a first set of results concerning the application of localization techniques to the chiral de Rham and structure sheaves of a toric variety $X$. We formulate a method to easily compute the partition function and the elliptic genus using an extension of the localization formula for the zero mode contributions. For that purpose we present in~\ref{sec:gaugetheory} a simple example showing how one can translate the information about the space into an abelian gauge theory. 
In~\ref{sec:two} we develop a general theory to describe these new models. It turns out that the natural setting are toric varieties. We can say that our gauge theory realizes the holomorphic quotient of a toric variety. (However, the formalism is not restricted to this case since it is easy to see how it can be extended to non-abelian gauge symmetries.) For this reason, we review in Section~\ref{sec:toric} the relevant facts about toric geometry which will be needed in this work. Section~\ref{sec:localization} concerns the localization formula. We give some arguments for deriving it from the well-known Hirzebruch-Riemann-Roch formula. The localization formula is translated into a path integral in the complex plane and the computation amounts to taking the residues at a certain set of poles. 

In the remainder of the text, we provide several examples to explain the computational technique and give an interpretation of the results. One of the main results is that the technique can be used for both compact and non-compact toric geometries. The information about the compactness is encoded only in the weights entering the definition of the integrand. The first set of examples are related to non-compact toric varieties in Section~\ref{sec:non-compact}. We apply our formula to orbifolds, their resolution in terms of a total space of a certain line bundle in Section~\ref{sec:orbifolds}. The latter is a member of a class of line bundles which we study in Section~\ref{sec:O_n}. Of course, we also consider the prevailing example of a non-compact Calabi-Yau manifold, the conifold, in Section~\ref{sec:conifold}. Compact examples are discussed in Section~\ref{sec:compact}. In particular, for $\mP^{1}$ and $\mP^{2}$ we can compare our results with those obtained by Borisov and Libgober. There are more interesting spaces which can be dealt with our formula, non-reduced schemes in Section~\ref{sec:Nonreducedscheme} and supermanifolds in Section~\ref{sec:superman}. A summary and further interesting topics are then given in Section~\ref{sec:conclusions}.

%%%%%%%%%%%%%%%%%%%%%%%%%%%%%%%%%%%%%%%%%%

\section{General theory}
\label{sec:two}

%%%%%%%%%%%%%%%%%%%%%%%%%%%%%%%%%%%%%%%%%%

\subsection{Non-linear $\beta\gamma$ systems and the chiral sheaves }
\label{sec:betagamma}

In this section we define the chiral algebras that are the objects of our interest. A chiral algebra is, in physicists' language, a set of fields ${\cal O}_{i} (z)$ depending holomorphically on the coordinate $z$, and closed under the operator product expansion: 
$$ {\cal O}_{i}(z)  {\cal O}_{j}(z') = \sum_{k} f_{i,j} {}^{k} (z-z') {\cal O}_{k} (z')$$ 
where the coefficients $f_{i,j}{}^{k}$ are holomorphic functions. As mentioned in the introduction, chiral algebras arise typically from the quantization of sigma-models in particular limits. In an ordinary sigma model, if it is conformally invariant, the set of fields will have a holomorphic and a antiholomorphic sector, and the two sectors can be decoupled by taking the infinite-volume limit. It is then legitimate to study each sector separately~\cite{Kapustin:2005pt}. If the sigma model has $N=2$ supersymmetry in one sector (say in the antiholomorphic), then one can perform the topological twist; the twisted algebra contains a BRST operator, and taking the BRST cohomology removes the antiholomorphic dependence from the spectrum of operators, hence reducing the problem again to a chiral sector. This is the half-twisted sigma model. The supersymmetric and non-supersymmetric case have very different features, and in the following we will discuss both. 

In~\cite{Malikov:1998dw},\cite{Malikov:1999ab} the authors showed that to every complex manifold $X$ one can canonically associate a sheaf of chiral (vertex) algebras, called the chiral de Rham complex $\cdr{X}$. This means that on each open subset $U \subset X$ the space of sections $\cdr{X}(U) = \Gamma(X,\cdr{X})$ is a chiral algebra, and when $U \subset V$ there are restriction maps $\Gamma(V,\cdr{X}) \to \Gamma(U, \cdr{X})$ that are algebra morphisms. Since one can always cover the manifold with charts that have the topology of $\mC^{n}$, it is enough to define the chiral algebra in this case, and then give the map that allows to identify the sections of $U_{i}$ and $U_{j}$ in the overlap $U_{i} \cap U_{j}$. To go from the theory defined on each single patch $U_{i}$ to the globally defined theory we need to prescribe a gluing procedure. The obstruction to do this is encoded in the \v Cech cohomology 
$\check{H}^2(X,\Omega^{2,cl}(X))$ of closed 2-forms on $X$~\cite{Gorbounov:1999ab},~\cite{Gorbounov:2000ab},~\cite{Gorbounov:2000ac}. 

The chiral algebra for $\mC^{n}$ is generated by a set of fields $\gamma^{i},\beta_{i}, c^{i}, b_{i}, i = 1, \dots, n$. The fields $\gamma,\beta$ are bosonic; the $\gamma$'s represent the coordinates $x^{i}$ of $\mathbb{C}^{n}$ and the $\beta$ are their conjugate momenta (from mathematical point of view the conjugate momenta are seen as the pull-back of tangent vectors living on the tangent bundle of the space $X$.) The fields $c,b$ are fermionic, and $c^{i}$ are in correspondence with the 1-forms $dx^{i}$. The OPEs are those of free fields: 
\begin{align}
  \label{freeOPE}
\g^{i}(z) \b_{j} (w) &= \frac{\delta^{i}_{j}}{z-w} \,, & c^{i}(z) b_{j} (w) &=  \frac{\delta^{i}_{j}}{z-w} \, . 
\end{align}
The chiral algebra contains an $N=2$ topological algebra:
\begin{eqnarray} 
  \label{N2gen}
T(z) &=&    :\b_{i} \partial \g^{i}:(z) - :b_{i} \partial c^{i}:(z)  \nonumber \\ 
Q(z) &=&    c ^{i} \b_{i} (z) \nonumber \\ 
G(z) &=&   b_{i} \g^{i} (z)  \\ 
J(z) &=&  :b_{i} c^{i}: (z) \nonumber    
\end{eqnarray}
Let us notice that the chiral de Rham complex is bigraded: $\cdr{X} = \oplus_{i,n} (\cdr{X})^{,i}_{(n)}$ where $i \in \mZ$ is the fermion number and $n \in \mZ_{+}$ is the conformal weight, measured by the OPE with the $U(1)$ current $J$  and the stress-energy tensor $T$, respectively. When a field of weight $h$ is expanded as a power series in $z$, $\phi(z) = \sum \phi_{n}/z^{n+h}$, the component $\phi_{n}$ is an operator of weight $-n$. If we restrict the complex to the zero modes (operators of weight zero), then it reduces to the ordinary de Rham complex $\Omega^*_X$, which explains the name. 

The crucial observation made in~\cite{Malikov:1998dw} is that, if we define the chiral algebra in a different patch using another set of fields $\tilde \g, \tilde \b, \tilde c, \tilde b$, with the OPEs as in (\ref{freeOPE}), it is consistent to identify the fields in the overlap according to the following rule: if $\tilde x^{i} = f^{i}(x^{j})$ are the transition functions for the coordinates in the two patches, with the inverse transformation 
$x^{i} = g^{i}(\tilde x)$, then 
\begin{eqnarray}
  \label{cdrglueing}
\tilde \g^{i}  & = & f^{i}(\g) \, ,  \nonumber \\
\tilde \b_{i}  & = &\frac{\partial g^{j}}{\partial \tilde\g^{i}} \left(f(\g)\right)\b_{j} + \frac{\partial^{2} g^{k}}{\partial \tilde\g^{i} \partial \tilde \g^{l}} \left(f(\g)\right)\frac{\partial f^{l}}{\partial\g^{r}} b_{k} c^{r} \,, \nonumber\\
\tilde c^{i} & =  &\frac{\partial f^{i}}{\partial \g^{j}} c^{j} \, ,  \nonumber \\
\tilde b_{i}  & = & \frac{\partial g^{j}}{\partial \tilde\g^{i}} \left(f(\g)\right) b_{j}
\end{eqnarray}
The fields $\g,c,b$ have the same transformation laws as the corresponding geometrical objects, but the second term in the transformation of $\beta$ is a quantum effect (in \cite{Nekrasov:2005wg} a complete analysis of this issue is presented and it is shown how this is related to a WZW action). If it were not for this term, the chiral de Rham sheaf would be the same as the complex: 
\begin{equation}
  \label{ellcomp}
  {\cal{ELL}}(X) (y,q) = \otimes_{k\geq 1} (\Lambda^{\bullet}_{-y q^{k-1}} T_X \otimes \Lambda^{\bullet}_{-y^{-1} q^{k}} T^{*}_X \otimes S^{\bullet}_{q^{k}} T_X \otimes S^{\bullet}_{q^{k}} T^{*}_X  ) \,.
\end{equation}
Here we use the notation $\Lambda^{\bullet}_{t} V = \sum_{i} t^{i} \Lambda^{i} V$ and $S^{\bullet}_{t} V = \sum_{i} t^{i} S^{i} V$ where $\Lambda^{i} V$ and $S^{i} V$ are the antisymmetric and symmetric tensor products of the vector bundle $V$, respectively.  $T_X$ is the holomorphic tangent bundle, and $T^*_X$ its dual. The reason for introducing the parameters $q,y$ will become apparent shortly. The holomorphic Euler character of this complex is a topological invariant of $X$ known as the elliptic genus. This is a function of two variables and can be computed with the Hirzebruch-Riemann-Roch formula  
\begin{equation}
  \label{eq:HRR}
  \Ell_{X}(y,q) = \int_{X} \chern({\cal{ELL}}(X) (y,q)) \todd(X) 
\end{equation}
where $\chern$ is the Chern character and $\todd$ is the Todd class~\cite{Hirzebruch:1992ab}. It is possible to show that the chiral de Rham complex has a filtration, and the associated graded complex is exactly the complex (\ref{ellcomp}) 
(see~\cite{Borisov:1998dw}, \cite{Borisov:1999ab}). A consequence of this fact is that the elliptic genus is also the Euler character of the chiral de Rham complex: 
\begin{equation}
  \label{eq:ellipticgenus}
\Ell_{X}(y,q) = {\mathrm Tr}_{{\cal F}_X}\left( (-y)^{F_0} q^{L_0}\right) = \sum_{i,p; n \geq 0} (-)^{i+p} y^{i} q^{n} \dim \,\check{H}^{p}(X, (\cdr{X})^{,i}_{(n)}) \, .
\end{equation}
We will often call such a sum of graded dimensions a character, denoted $\mathrm{Ch}$. We can then write 
\begin{equation} 
  \Ell_X(y,q) = \sum_{p} (-)^{p} \, \mathrm{Ch}\left( H^{p} (\cdr{X} ) \right) 
\end{equation}
It is useful to remind the reader that among the spaces $\check{H}^{p}(X, (\cdr{X})^{,i}_{(n)})$ there are some relations due to the Poincar\'e duality~\cite{Malikov:1999ac} (in the case of compact spaces of total dimension $d$) of the form
\begin{equation}
  \label{eq:poinA}
  (\check{H}^{p}(X, (\cdr{X})^{,i}_{(n)}))^{*} = \check{H}^{d-p}(X, (\cdr{X})^{,d-i}_{(n)})\,.
\end{equation}
This formula is useful for comparing the cohomological result with the partition function computations. 

As will be explained in more details later, if there is some group acting holomorphically on $X$, it is possible to substitute the character with the equivariant character, depending on more variables $t$ associated to the group action. Then~
\begin{equation}
  \label{eq:ellipticgenus}
\Ell_{X}(t,y,q) = {\mathrm Tr}_{{\cal F}_X}\left( (-y)^{F_0} t^{K_0} q^{L_0}\right) = \sum_{i,p; n,m \geq 0} (-)^{i+p} y^{i} t^{m} q^{n} \dim \,\check{H}^{p}_{m}(X, (\cdr{X})^{,i}_{(n)}) \, .
\end{equation}
which was mentioned in the introduction. Among the operators of the $N=2$ algebra, $T$ and $G$ are invariant under the coordinate transformations and so they are global sections of the chiral sheaf; the other operators $Q$ and $J$ are not invariant unless the manifold is Calabi-Yau. 

The discussion can be repeated for the non-supersymmetric version, but only to some extent. In this case we only have the $\b\g$ system.  Without the fermions, one cannot write down the change of coordinate on $\beta$ as in~(\ref{cdrglueing}). The only possibility is to try an Ansatz of the form 
\begin{equation}
\tilde \b_{i} = \frac{\partial g^{j}}{\partial \tilde\g^{i}} \left(f(\gamma)\right)\b_{j}  + B_{ij}(\gamma) \partial \gamma^{j} \, . 
\end{equation}
One has to find a suitable $B_{ij, \a\b}$ in each intersection $U_{\a} \cap U_{\b}$; it can be chosen to be antisymmetric in $ij$ so it can be identified with a two-form $B = B_{ij} d\g^{i} \wedge d\g^{j}$. But there is a consistency condition on triple overlaps: 
$B_{\a,\b}+B_{\b,\g}+B_{\g,\a} = 0$. There is a potential obstruction to satisfy this equation, measured by the \v Cech cohomology group ${\check{H}}^{2}(X, \Omega_{X}^{2,cl})$ of closed 2-forms on $X$~\cite{Gorbounov:1999ab}, \cite{Witten:2005px}, \cite{Nekrasov:2005wg}. In terms of de Rham cohomology this is the first Pontrjagin class of the manifold. Only when $p_{1}(X) = 0$ it is possible to find a consistent set of gluing rules and to have a globally well-defined sheaf, which is then called the chiral structure sheaf $\css{X}$.  There is also another anomaly that can only be seen when studying the $\beta\gamma$ system on a higher genus Riemann surface $\Sigma$. It is proportional to $c_{1}(X) c_{1}(\Sigma)$, but in this paper we will work on the torus so we do not need to discuss it (see \cite{Witten:2005px}, \cite{Nekrasov:2005wg}). 
Finally, just as for the chiral de Rham complex, one can define the (equivariant) Euler character of $\css{X}$:
\begin{equation}\label{eq:eulerssh}
  Z_{X} (t,q) = \sum_{p} (-1)^p\, \mathrm{Ch} \left( H^{p} (X, \css{X})\right)\,. 
\end{equation}%
This function contains some geometric information about $X$, but it is not a topological invariant, in particular it can depend on the complex structure moduli of $X$. In the rest of the paper we will be concerned with the computation of $\Ell_{X}$ and $Z_{X}$ for several classes of manifolds.  

We can also consider the situation in which we have $\beta\gamma$ and $bc$ systems that are not related by supersymmetry. In this case, the chiral algebra is generated by fields $\gamma^i,\,\beta_i$, $i=1,\dots,n$ and $c^a,\,b_a$, $a=1,\dots,r$. Since there is no supersymmetry, $n$ and $r$ do not necessarily need to be the same. Geometrically, this corresponds to having a complex manifold $X$ of dimension $n$ together with a holomorphic vector bundle $E \to X$ of rank $r$. An analogous argument as in the case of the chiral structure sheaf shows that the obstruction to have a globally defined chiral sheaf is that $p_1(E) = p_1(X)$~\cite{Gorbounov:1999ab}, \cite{Tan:2006qt}. In the special cases $E=\Omega^1_X$ and $E=\cO_X$, this reduces to the chiral de Rham complex and the chiral structure sheaf, respectively.

%%%%%%%%%%%%%%%%%%%%%%%%%%%%%%%%%%%%%%%%%%

\subsection{The $\beta\gamma$ system as a gauge theory}
\label{sec:gaugetheory}

The main motivation for the next two sections is to illustrate using two very simple examples how localization can be used to compute the partition function and elliptic genus. The general formula will be given in~\ref{sec:localization}. 

%%%%%%%%%%%%%%%%%%%%%%%%%%%%%%%%%%%%%%%%%%

\subsubsection{The bosonic case}
\label{sec:bosonic}

We consider the simple $\gamma=AB$ model where the field  $\gamma: \Sigma \to X$ (where $\Sigma$ is a Riemann surface, $X$ is a complex manifold which we take for the moment to be $\mC$ ) is split into two fields $A$ and $B$. The $\gamma$-model and the $AB$-model are equivalent if we impose the gauge symmetry  $A \rightarrow \Lambda A$ and $B \rightarrow \Lambda^{-1} B$ (all fields $A,B$ and $\Lambda$ are  $\mC^{\infty}(\Sigma)$ ). We introduce the conjugate momenta $\beta,\beta_{A},\beta_{B}$ which are elements of $\Omega^{(1,0)}_{\Sigma}$. The action of the model written in the coordinates $\gamma$ or $A,B$ is 
\begin{equation}
  \label{eq:acA}
S_{AB} =\int d^{2}z \beta \bar\p \gamma = 
\int d^{2}z \Big( \beta_{A} \bar \p A + \beta_{B} \bar\p B\Big)\,.     
\end{equation}
The equivalence between the two action gives us relations $\beta_{A} = \beta B$ and $\beta_{B} = \beta A$. All fields are chiral and they have the following elementary OPE's
\begin{align}
  \label{eq:acB}
  \gamma(z) \beta(w) &\sim \frac{1}{ z-w}\,, & 
  A(z) \beta_{A}(w) &\sim \frac{1}{ z-w}\,, & 
  B(z) \beta_{B}(w) \sim \frac{1}{ z-w}\,.  
\end{align}
Notice that $\beta_{A}(z) \beta_{B}(w) \rightarrow 0$ as it can be verified by using~(\ref{eq:acB}) and $\beta A(z) B(w) \rightarrow \beta(z) (AB)(w) = -1/(z-w)$. Attention must be paid to manipulate with the inverse relations $\beta = \beta_{A} B^{-1} = \beta_{B} A^{-1}$. 

Performing the variation $A \rightarrow \Lambda A$ and $B \rightarrow \Lambda^{-1} B$, we also need that $\beta_{A} \rightarrow \Lambda^{-1} \beta_{A}$ and 
$\beta_{B} \rightarrow \Lambda \beta_{B}$ and the constraint 
\begin{equation}
  \label{eq:acC}
J_{z} = :\beta_{A} A - \beta_{B} B: =0\,.  
\end{equation}
This constraint directly follows from the definition of $\beta_{A}$ and $\beta_{B}$. It is needed in order to make the counting of the degree of freedom work: the gauge symmetry removes one degree of freedom among $A$ and $B$, the constraint~(\ref{eq:acC}) fixes one of conjugate momenta in term of the other.  The current $J_{z}$ is gauge invariant. 
The action is also invariant under the rigid symmetry generated by the zero mode of $J_{0} = \oint dz J_{z}$ which scales the fields $A$ and $B$. However, the action has another rigid symmetry generated by $K_{0} = \oint dz (\beta_{A} A + \beta_{B} B)$. Combining the two rigid symmetries $J_{A} = J_{0} + K_{0}$ and $J_{B} = J_{0} - K_{0}$, we see the action is invariant under the separate rescaling of the fields $A$ and $B$. 

{We have denoted by $J_{0}$ a generic charge associated to some rigid symmetry. The definition of a character or elliptic genus in this context can be traced back to~\cite{Witten:1986bf}, \cite{Witten:1987cg}. A convenient way to formulate a character in quantum field theory is to introduce a gauge field $A_{\bar z}$ to render the rigid symmetry generated by $J_{0}$ a local gauge symmetry. Then, after fixing the gauge $A_{\bar z} = (2 \pi i w)$ where $w$ is the constant vector of the torus (monodromy), we see that the action reduces to $2 \pi i J_{0}$. Finally, we identify $t = e^{2 \pi i w}$.} Obviously, a correct gauge fixing procedure boils down 
to introducing some ghost fields and they are indeed necessary to cancel unwanted degrees of freedom. In the introduction we have mentioned that the the world sheet instantons do not enter the computation we are performing. We can motivate this further by observing that only one component of the gauge field is entering the theory, hence there is no kinetic term in the action for the gauge field neither a $\theta$ term. 

Now, we would like to show that the system $AB$ is equivalent to the system of a single free field $\gamma$ if we correctly implement the gauge symmetry at the level of the partition function. We can view the free field $\gamma$  as living on $\mC^{2}/ \mC^*$ where the $\mC^*$ is the (complexified) gauge symmetry implemented by $\Lambda$. Notice that the gauge parameter is a local holomorphic parameter and therefore it can be decomposed into modes 
\begin{equation}
  \label{eq:exaA}
  \Lambda(z) = \sum_{n=-\infty}^{\infty}  \Lambda_{n} z^{n}\,,
\end{equation}
where the zero mode $\Lambda_{0}$ is the overall rescaling of the two field $A$ and $B$ generated by $J_{0}$. After the 
gauge fixing the ghost fields replace the gauge parameters.

Now, the (character) partition function
\begin{equation}
  \label{eq:acDA}
  Z_{\gamma}(t|q) = {\rm Tr}_{\cal H}\left(t^{J_{0}} q^{L_{0}} \right)   
\end{equation}
for the field $\gamma$ (and its conjugate $\beta$) is simply given by 
\begin{equation}
  \label{eq:acD}
  Z_{\gamma}(t|q) = \frac{1}{1-t}\prod_{n=1}^{\infty}\frac{1}{ (1- t q^{n})(1 - t^{-1} q^{n})} =  -i \sqrt{t} q^{-1/12} \frac{\eta(q) }{ \theta_1(t|q) }\,,  
\end{equation}
where $t$ is associated to the rescaling of the field $\gamma$. The space ${\cal H}$ is represented by the Fock space of the holomorphic polynomials generated by $\gamma$ and $\beta$ modes. The definition of $\eta(q)$ and $\theta_{1}(t|q)$ are the Dedekind function and the first Jacobi modular form with character. The first factor in the denominator represents the zero mode part, the second and third represent the non-zero modes of $\gamma$ and $\beta$, respectively (we would like to remind the reader that a basis for a the chiral algebra on each single 
patch is given be the operators $\p^{p} \gamma$ and $\p^{p} \beta$ for any $p \geq 0$.)

Note that the partition function is divergent for $t=1$. This is due to the presence of non-normalizable zero modes for the non-compact space $\mC$. The presence of $t$ equivariantly reduces the space to a single point with on trivial volume and the zero modes are represented by $1/(1-t)$. Using the linearized group action $t = e^{2 \pi i \epsilon}$, we see that in the limit $\epsilon \rightarrow 0$ we get the regularized volume of a point (see e.g.~\cite{Moore:1997dj}). The cohomology corresponds to the equivariant cohomology and the equivariance is measured by $t$.

We would like to reproduce the same partition function starting from the system $AB$. Following~\cite{Nekrasov:2004vw} we can show that the zero modes $Z^{0}_{AB}(t|q)$ coincide by implementing the gauge symmetry 
\begin{equation}
  \label{eq:gaA}
  Z^{0}_{AB}(t_{A}, t_{B}|q) = \oint \frac{d\Lambda_0 }{ \Lambda_0} \frac{1}{ (1 - t_{A} \Lambda_0)( 1- t_{B}\Lambda_0^{-1})}
\end{equation}
where ${d\Lambda_0/ \Lambda_0}$ is the measure of the orbit space, $\Lambda_0$ is the parameter of the gauge symmetry\footnote{To be precise we would have put the measure $dh \prod_{\alpha \in \Delta} (h^{\alpha/2} - h^{- \alpha/2})/Vol(T_{H})$ where $h$ are the group representative of the maximal torus $T_{G}$ of the Lie group $G$. $\Delta$ is the set of positive roots. For more details we refer to~\cite{Nekrasov:2004vw} and for a complete discussion to~\cite{Pressley:1988qk}. For the moment we have the simplification that all torus actions are abelian and therefore the measure reduces to $\prod_{i=1}^{n} d\Lambda_{i} / \Lambda_{i}$ where $\Lambda_{i}$ are the gauge parameters of the $\mC^{*}$ actions. The volume is $(2 \pi i)^{n}$.} and the two fields $A$ and $B$ are rescaled by $t_{A}$ and $t_{B}$ in the partition function. We observe that the integrand has a single pole at $\Lambda_0 = t_{B}$ and by the residue computation we get
\begin{equation}
  \label{eq:gaB}
  Z^{0}_{AB}(t_{A}, t_{B}|q) = \frac{1 }{ (1 - t_{A} t_{B})}
\end{equation}
which concides with the zero mode part of~(\ref{eq:acD}) if $t = t_{A} t_{B}$. To extend this result to the complete partition function~(\ref{eq:acD}) we need to implement the remaining gauge symmetries adding the contribution of the gauge modes as a numerator $\prod_{k=1}^{\infty} (1 - q^{n})^{2}$ (this product involves only the massive modes and the exponent 2 is due to positive and negative modes of the expansion~(\ref{eq:exaA}) )
\begin{eqnarray}
  \label{eq:gaC}
  Z_{AB}(t_{A}, t_{B}|q) &=& \oint \frac{d\Lambda_0 }{ \Lambda_0} 
  \frac{1}{ (1 - t_{A} \Lambda_0)( 1- t_{B}\Lambda_0^{-1})}\times\cr
  && \phantom{\oint \frac{d\Lambda_0 }{ \Lambda_0}}\prod_{n=1}^{\infty} 
  \frac{(1- q^{n})^{2} }{ 
  (1 - t_{A} \Lambda_0 \, q^{n})(1 - \frac{1}{ t_{A} \Lambda_0} q^{n})
  (1 - \frac{t_{B} }{ \Lambda_0} q^{n})
  (1 - \frac{\Lambda_0 }{ t_{B}} q^{n}) }\cr
  &=&\frac{1 }{ (1 - t_{A} t_{B})} \prod_{n=1}^{\infty}\frac{1 }{ (1- t_{A} t_{B} q^{n})(1 - \frac{1}{ t_{A} t_{B}} q^{n}) } \cr
  &=& Z_{\gamma}(t_{A}t_{B}|q)\,,
\end{eqnarray}
where we have assumed that the only pole that contributes is at $\Lambda_0= t_{B}$. Notice that by inserting the value $\Lambda_0=t_{B}$ into the integrand, the contribution of $B$ and $p_{B}$ exactly cancels against the gauge symmetry modes $\prod_{n>0} (1- q^{n})^{2}$. In order to understand how the gauge symmetry works we notice that gauge parameters $\Lambda_{0}, \Lambda_{-1}, \Lambda_{-2}, \dots$ 
transform the modes of the field $A$ as follows 
\begin{align}
  \label{eq:gaD}
  \delta A_{0} &= \Lambda_{0} A_{0}\,, &
  \delta \left(\frac{A_{-1}}{ A_{0}}\right) &= \Lambda_{-1}\,, &
  \delta \left({A_{-2} }{ A_{0}} - \frac{1}{ 2}\frac{ A^{2}_{-1} }{ A^{2}_{0}}\right)   &= \Lambda_{-2}\,, & \dots
\end{align}
We see from the first equation that the zero mode of the gauge parameter $\Lambda_{0}$ implements a ``true'' gauge symmetry rescaling locally the zero mode of $A$ and $B$. The remaining equations implement a shift of the massive modes and remove two of each at each level. Notice furthermore that the shifts are scale invariant and this is reflected in the numerator $\prod_{n>0} (1- q^{n})^{2}$ which represent a set of scale invariant fermionic modes. 

To complete the discussion, we need to clarify the discussion on which poles have to be taken into account. Notice that in principle there are two sets of poles $\Lambda_{0,n} = q^{n}/ t_{A}$ and $\Lambda_{0,n} = t_{B} q^{n}$. We will argue in the following section how to prescribe the set of poles to be taken into account from the zero mode part of the partition fuction. After this is done, we can use the same poles to compute the total partition function by taking the residue at these poles.

%%%%%%%%%%%%%%%%%%%%%%%%%%%%%%%%%%%%%%%%%%

\subsubsection{The supersymmetric case}
\label{sec:supersymmetric}

We add fermions to the previous model. It is convenient to introduce a single anticommuting field $c$ and its conjugate $b$ such that the action 
\begin{equation}
  \label{eq:ferdecA}
  S = \int d^{2}z \left(\beta \bar\partial \gamma + b \bar\partial c\right)   
\end{equation}
is invariant under the supersymmetry
\begin{align}
  \label{eq:ferdecB}
  \delta \gamma &= c\,, & \delta c &= \p \gamma\,, & \delta b &= - \beta\,, & \delta \beta &= \p b\,.
\end{align}
Decomposing the field $\gamma$ as above, we also need a decomposition 
for the fermions: 
\begin{equation}
  \label{eq:ferdecC}
  \gamma = AB\,, \qquad c = c_{A} B + A c_{B}
\end{equation}
where we have introduced $c_{A}, c_{B}$ and their conjugate momenta $b_{A}, b_{B}$. We also have doubled the anticommuting fields and we need a new gauge symmetry to remove them. Notice that $\delta (AB) = c_{A} B + A c_{B}$ if $\delta A = c_{A}, \delta c_{A} = \p A$ and $\delta B = c_{B}, \delta c_{B} = \p B$. So, the supersymmetry for the composing fields $c_{A},c_{B}, A, B$  is needed in order to reproduce the supersymmetry of the original theory. In addition, because of the supersymmetry the fields $c_{A}$ and $c_{B}$ have the same scaling behaviour as $A$ and $B$, respectively. This can be seen by inserting the decomposition~(\ref{eq:ferdecC}) into~(\ref{eq:ferdecA}), and identifying
\begin{align}
  \label{eq:ferdecD}
  \beta_{A} &= \beta A + b c_{A}\,, & \beta_{B} &= \beta B + b c_{B}\,, & b_{A} &= b B\,, & b_{B} &= b A\,.
\end{align}
as the conjugated momenta. From these definitions, one gets the relations 
\begin{equation}
  \label{eq:ferdecE}
  :\beta_{A} A + b_{A} c_{A} - \beta_{B} B - b_{B} c_{B}: =0\,, \qquad :b_{A} A - b_{B} B: = 0\,.
\end{equation}
Only in the first relation, the normal ordering is needed. Note that $\delta(:b_{A} A - b_{B} B:) = :\beta_{A} A + b_{A} c_{A} - \beta_{B} B - b_{B} c_{B}:$. The second constraint implies the new gauge symmetry 
\begin{align}
  \label{eq:ferdecF}
  \Delta A &= \Delta B =0\,, & \Delta c_{A} &=\epsilon A\,, & \Delta c_{B} &= - \epsilon B\,,\cr 
  \Delta \beta_{A} &= - \epsilon b_{A}\,, & \Delta \beta_{B} &= - \epsilon b_{B}\,, & \Delta b_{A} &= \Delta b_{B} =0\,.
\end{align}
where $\epsilon$ is an anticommuting gauge parameter. The relations~(\ref{eq:ferdecC}) are invariant under the new gauge symmetry. This gauge symmetry is needed in order to reduce the number of the fermionic fields $c_{A}, c_{B}$ to just the original one $c$. The constraints~(\ref{eq:ferdecE}) are needed in order to reduce the number of independent conjugate momenta $b_{A}, b_{B}, \beta_{A}, \beta_{B}$ to the original $\beta, b$. Finally, we can define a new gauge invariant charge $Q = \oint (b_{A} c_{A} + b_{B} c_{B})$ which counts the fermion number and we denote its parameter by $y$. 

Repeating the computation of the previous section with fermions yields: 
\begin{eqnarray}
  \label{eq:ferdecG}
  Z_{\gamma +c}(y, t_{A}, t_{B}|q) &=& 
  \oint \frac{d \Lambda_{0} }{ (1- y) \Lambda_{0}} 
  \frac{(1 - \Lambda_{0} y t_{A}) (1 - \Lambda^{-1}_{0} y t_{B}) }{ 
  (1 - \Lambda_{0} t_{A}) (1 - \Lambda^{-1}_{0} y t_{B})} \times\cr
  &&\phantom{\oint \frac{d \Lambda_{0} }{ (1- y) \Lambda_{0}}}\prod_{k\geq 1} \frac{(1 - \Lambda_{0} y t_{A} q^{k}) (1 - \frac{1}{ \Lambda_{0} y t_{A}} q^{k}) (1 -  \frac{y t_{B} }{ \Lambda_{0}} q^{k}) (1 - \frac{\Lambda_{0} }{ y t_{B}} q^{k})}{ (1 - \Lambda_{0} t_{A} q^{k}) (1 - \frac{1}{ \Lambda_{0} t_{A}} q^{k})(1 -  \frac{t_{B} }{ \Lambda_{0}} q^{k})  (1 - \frac{\Lambda_{0}}{ t_{B}} q^{k})} \times\cr
  & & \phantom{\oint \frac{d \Lambda_{0} }{ (1- y) \Lambda_{0}}}\prod_{k\geq1}\frac { (1 - q^{k})^{2} }{ (1- y q^{k})(1- y^{-1} q^{k})} \cr
  &=& \frac{(1 - y t_{A} t_{B})}{ (1- t_{A} t_{B})} \prod_{k\geq 1} \frac{(1 - y t_{A} t_{B} \, q^{k}) (1 - \frac{1}{ y t_{A} t_{B}} \, q^{k})}{ (1- t_{A} t_{B} \, q^{k}) (1- \frac{1}{ t_{A} t_{B}}\,  q^{k})} \cr
 &=& \sqrt{y} \frac{\theta_{1}( y\,  t_{A} t_{B}|q)}{ \theta_{1}(t_{A}t_{B}|q)}\,.
\end{eqnarray}
The last expression coincides with the partition function for a free boson with charge $t = t_{A} t_{B}$ and for a free fermion with charge $y t_{A} t_{B}$ which is the correct result. Notice that the denominator $1-y$ in the integrand represents the contribution of the zero mode of the supergauge symmetry~(\ref{eq:ferdecF}), namely the ghost field associated to the gauge parameter $\epsilon$.

%%%%%%%%%%%%%%%%%%%%%%%%%%%%%%%%%%%%%%%%%%

\subsection{Toric Geometry}
\label{sec:toric}

What we have done in the previous section from the geometrical point of view is to view $\mC$ as a toric variety $\mC^2/\mC^*$. This suggests that we can generalize the result to arbitary toric varieties. In this section, we therefore summarize some of the basic facts about toric geometry. For a very nice short introduction see e.g.~\cite{Kreuzer:2006ax} while further details can be found in e.g.~\cite{Fulton:1993ab}. Then we discuss how toric varieties appear in the context of localization. 

An intuitive way to describe a toric variety is to say that it is a variety that contains an algebraic torus $\mT=\left(\mC^*\right)^n$ as a dense open subset such that the natural action of $\mT$ on itself extends to an action of $\mT$ on $X$.

In more practical terms, an $n$--dimensional toric variety takes the form
\begin{equation}
  \label{eq:XSigma}
  X_\Sigma = \left(\mC^d \setminus F_\Sigma \right) / \left(\mC^*\right)^r,
\end{equation}
where $n=d-r$. The group $\left(\mC^*\right)^r$ acts by coordinatewise multiplication. $F_\Sigma$ is a subset that is being fixed under the action of a continuous subgroup of $\left(\mC^*\right)^r$ and will be described in detail below. The action of $\left(\mC^*\right)^r$ is encoded in a structure, called the fan $\Sigma$. This is a finite collection of strongly convex rational polyhedral cones in a lattice with the property that each face of a cone in $\Sigma$ is itself a cone in $\Sigma$ and the intersection of two cones in $\Sigma$ is a face of each. 

The space in which $\Sigma$ lives can be obtained as follows: Consider the algebraic torus $\mT=\left(\mC^*\right)^n$ with coordinates $(t_1,\dots,t_n)$. There are two natural lattices associated to $\mT$. The first one is the character group $M=\left\{\chi:\mT \to \mC^*\right\}$ which can be identified with a lattice $M\cong \mZ^n$, where $m\in M$ corresponds to the character $\chi^m(t_1,\dots,t_n) = t_1^{m_1}\dots t_n^{m_n} \equiv t^m$. The second one can be identified with the group of algebraic one--parameter subgroups $N\cong \left\{ \lambda : \mC^* \to \mT \right\}$ where $v \in N$ corresponds to the group homomorphism $\lambda^u(\tau) = \left(\tau^{v_1},\dots,\tau^{v_n}\right) \in \mT$ for $\tau \in \mC^*$ \footnote{In order to avoid confusion with the dimension $n$, we write in this section $v_i$ instead of $n_i$ for vectors in the lattice $N$. In the remaining sections, we will write $n_i$.}. These two lattices are dual to each other in the following way: The composition $(\chi \circ \lambda)(\tau) = \tau^{\langle \chi, \lambda \rangle}$ defines a canonical pairing $\langle \chi^m, \lambda^v \rangle = m \cdot v$. The characters $\chi^m$ for $m\in M$ can be regarded as holomorphic functions on $\mT$ and hence as rational functions on the toric variety $X$. This duality has a counterpart in the localization as we will see later on.

Once we have the lattice $N$, we can describe the fan $\Sigma$ by its generators, i.e. by its 1--dimensional cones which are lattice vectors $v_i\in N,\; i=1,\dots, d$. It can be shown that they are in 1--to--1 correspondence with the $\mT$--invariant divisors $D_i$ on $X_\Sigma$. If we locally write the equation of the divisor as $D_i=\{z_i = 0\}$, with $z_i$ being a section of some line bundle, we can regard the $\{z_j\}$ as global homogeneous coordinates. In terms of these coordinates we can define the set $F_{\Sigma}$ in~(\ref{eq:XSigma}). A subset of the coordinates $\{z_i\}$ is allowed to vanish simultaneously if and only if there is a cone $\sigma \in \Sigma$ containing all of the vectors $v_i$. The set $F_{\Sigma}$ is therefore the union of the sets $F_I = \{(z_1,\dots,z_d)| z_i=0\, \forall i \in I\}$ for which there is no cone $\sigma \in \Sigma$ such that $v_i \subseteq \sigma$ for all $i \in I$. Minimal index sets $I$ with this property are called primitive collections~\cite{Batyrev:1991ab}. 

Having $d$ generators $v_i$ in a lattice of dimensions $n=d-r$ obviously means that we have $r$ linear relations of the form 
\begin{equation}
  \label{eq:linrels}
  \sum_{i=1}^d Q_{a,i} v_i = 0, \qquad a = 1,\dots,r\,.
\end{equation}
In terms of the gauge theory of the previous section, the $Q_{a,i}$ are the charges of the composing fields $A_i$ of $\gamma$ under the compact part of the $a$th factor of gauge group $\left(\mC^*\right)^r$. It is convenient to collect these data in a matrix of the form
\begin{equation}
  \label{eq:PQ}
  \left(
  \begin{array}{c|c}
    v_1 & Q_1 \\
    \dots & \dots \\
    v_d & Q_d 
  \end{array}
  \right)\,.
\end{equation}
The charges $Q_{a,i}$ in~(\ref{eq:linrels}) are only defined up to linear combinations and scalings of linear relations. To fix this ambiguity recall the notion of a primitive collection above. To each primitive collection $I$ there is an associated linear relation labeled by $a(I)$. A convenient set of basis vectors for the lattice of linear relations consists of a subset of $r$ of the primitive relations. We will use this basis to define our examples. Furthermore, this notion allows us to decompose the index set $I$ into $I=I_{+} \cup I_0 \cup I_{-}$ with~\cite{Batyrev:1991ab}
\begin{align}
  \label{eq:Iplusminus}
  I_{+} &= \{i \in I\,|\, Q_{a(I),i} > 0\}\,, &   I_{0} &= \{i \in I\,|\, Q_{a(I),i} = 0\}\,, &   I_{-} &= \{i \in I\,|\, Q_{a(I),i} < 0\}\,.
\end{align}
In fact, the minimality condition translates to $Q_{a(I_+),i} = 1$. While these subsets are related to coherent triangulations of the fan $\Sigma$ classified by the secondary polytope, here they will be relevant for the localization in the next subsection.

An alternative approach to define a toric variety $X_{\Sigma}$ is by taking a covering of $X_{\Sigma}$ by open affine patches $U_\sigma$ for maximal cones $\sigma \in \Sigma$. One way to define them is by taking the intersection of all complements $X_\Sigma \setminus D_i$ for $v_i \not\in \sigma$. Another way is to define $U_\sigma$ by taking the spectrum of maximal ideals of the ring $\mC[\sigma\spcheck \cap M]$. Here $\sigma\spcheck \in M$ is the cone dual to the cone $\sigma \in N$. In either case, $X_\Sigma$ is constructed by gluing the affine patches $U_\sigma,\,U_\tau$ along their intersection $U_\sigma \cap U_\tau = U_{\sigma \cap \tau}$. The cone $\sigma \cap \tau$ has dimension $n-1$. We can continue intersecting more patches and express the overlap of $k+1$ affine patches in terms of a cone of dimension $n-k$, until we end up with $U_{\{0\}} = \mT$.

Before we proceed with the localization we would like to remind the reader of two of the most important facts about toric varieties: First, a toric variety $X_\Sigma$ is compact if and only if the fan $\Sigma$ is complete, i.e. if the set of all cones covers $N\otimes\mR$. Second, a toric variety is non--singular if and only if all cones are simplicial and basic, i.e. if all cones $\sigma \in \Sigma$ are generated by a subset of a lattice basis of $N$.

%%%%%%%%%%%%%%%%%%%%%%%%%%%%%%%%%%%%%%%%%%

\subsection{Localization}
\label{sec:localization}

Our goal is to compute the partition function $Z_{X_\Sigma}(t,y|q)$ for toric varieties $X_\Sigma$. There are two ways to do this which are in some sense dual to each other. One way is to view $X_{\Sigma}$ as being cut out by some equations in a higher dimensional space, i.e. a $\mC^K$ in the non--compact case, or a $\mP^K$ in the compact case, for some large $K$. The equations are given in terms of $\mT$--invariant monomials, i.e. by the characters $\chi^m,\, m\in M$. Physically, this means that we impose a set of constraints on the theory. Constraints are notoriously difficult to deal with. For the zero mode part of the partition function for toric varieties, we can at least count the monomials by writing down the corresponding Poincar\'e polynomial. For the massive part, however, it is unclear how to proceed.\footnote{There have been a few attempts 
to pursue the computation of the partition function for the massive modes by 
counting directly the states appearing in the cohomology \cite{Grassi:2005jz}. 
One of the major problem is the absence of specific patterns that help reconstructing the counting at each level.}

The other way is to view $X_{\Sigma}$ as the quotient of a symplectic reduction. (For subtleties in the difference between the symplectic quotient and the holomorphic quotient~(\ref{eq:XSigma}) see e.g.~\cite{Kreuzer:2006ax} and references therein.) Here, we use gauge symmetries to reduce $\mC^K$ or $\mP^K$ to obtain $X_{\Sigma}$ via the moment map. The moment map is determined by the charges $Q_a$ of the gauge group, i.e. by the linear relations between the vectors in the lattice $N$. In other words, here we are working with the lattice dual to $M$. The latter contained the information in the approach using constraints. It is curious to note that the duality between constraints and gauge symmetries manifests itself here as the duality between the lattices $M$ and $N$, as described in the previous section.

In the set--up of symplectic reduction there is a powerful tool to compute the partition function, known as localization. For a review and further details see e.g.~\cite{Szabo:1996md}. Whenever there is an action by a Lie group $G$ on a manifold $X$, generated by a vector field $V$, it is possible to define the equivariant cohomology; this coincides with the cohomology of the quotient $X/G$ if the action is free. In the case with fixed points the easiest definition is by using the Cartan model: one considers the complex $\Omega^{\bullet} (X) \otimes  S( \mathfrak{g} ^{*})$, that is differential forms on $X$ with coefficients polynomials in the dual Lie algebra, and the restricted ``basic'' complex $\Omega_{G}$ of forms that are $G$-invariant,  where $G$ acts on $\mathfrak{g} ^{*}$ via the coadjoint action. 
The differential is modified: $d_{V} = d - \phi \iota_{V}$, where $\phi$ is a generator of $\mathfrak{g}^{*}$ and has degree 2, so that the differential is homogeneous of degree 1. It satisfies $d_{V}^{2} = {\cal L}_{V}$, and it is a differential on $\Omega_{G}$. The equivariant cohomology is now defined as 
$$H^{*}_{G} (X) = H^{*} (d_{V}, \Omega^{\bullet}_{G}) \,.$$
Any closed differential form $\alpha$ can be lifted to an equivariantly closed form $\tilde \alpha = \sum_{i} \phi^{i} \alpha_{i}$ where $\alpha_{0} = \alpha$ and the other terms, of lower form-degree, are found by solving $d \alpha_{i} = \iota_{V} \alpha_{i+1}$. In particular the characteristic classes of vector bundles can be extended to equivariant characteristic classes. 

The most important fact about equivariant cohomology is the localization formula: if the action of $G$ is not free, it has fixed points. Let us assume for simplicity that $X^{G}$, the fixed set of the action, contains only isolated fixed points, as this is always the case for toric varieties. Then for an equivariant cohomology class $\alpha$, 
\begin{equation}
\label{localization}
\int_{X} \alpha = \sum_{p \in X^{G}} \frac{\iota^{*}_{p} \alpha}{e(T_{p}X)}
\end{equation}
where $e$ in the denominator is the equivariant Euler class, and on the r.h.s. one takes the component of form-degree 0. Combining the Hirzebruch-Riemann-Roch formula with the localization formula, taking for $\alpha$ the equivariant Chern character of the complex~(\ref{ellcomp}), yields a formula for the elliptic genus of a toric variety $X_\Sigma$ in terms of the fixed points of the torus action. 

% Let a Lie group $G$ act on a symplectic manifold $(M,omega)$. A emph{moment map} is a map $Phi colon M to g^*$ to the dual of the Lie algebra such that the $G$ action is generated by the Hamiltonian vector fields of the components of $Phi$. The emph{symplectic quotient} is $Phiinv(0) / G$. emph{Localization} formulas express global invariants of $M$ in terms of local data at the fixed point set of an abelian subgroup of $G$. When $G$ is a torus of half the dimension of $M$ and $M$ is compact, $(M,omega,Phi)$ is a emph{toric manifold}.

We propose to extend the localization formula for the zero modes to the massive modes as follows. We define the following expressions
\begin{align}
t^{v_i} &= \prod_{k=1}^{n} t_{k}^{v_{i,k}}\,, &
z^{Q_i} &= \prod_{a=1}^{r} z_{a}^{Q_{i,a}}\,, 
\end{align}
where $ t_{k}$ with $k=1,\dots,n$ are the scaling parameters of the toric variety $X_\Sigma$, i.e. the coordinates of the torus $\mT$, and $z_{a}$ with $a=1,\dots,r$ are the gauge parameters of the symmetry $\left(\mC^{*}\right)^{r}$. Then, we can write the complete localization formula as follows
\begin{equation}
\label{eq:zero}
  \Ell_{X}(y,t|q) = \frac{1}{G(y|q)^r} \left(\prod_{a=1}^{r} \oint 
\frac{dz_{a}}{z_{a}}\right) \prod_{m=1}^{\infty}\prod_{i=1}^{d} 
\frac{ \left(1- y\, t^{v_i}\, z^{Q_i} \,q^{m-1} \right) 
\left(1- y^{-1}\, t^{-v_i}\, z^{-Q_i} \,q^{m} \right) }{
\left(1- t^{v_i}\, z^{Q_i} \,q^{m-1}\right) 
\left(1-  t^{-v_i}\, z^{-Q_k} \,q^{m}\right ) }
\end{equation}
where 
\begin{equation}
G(y|q) = \prod_{m=1}^{\infty} \frac{(1- y q^{m-1})(1- y^{-1} q^{m})}{(1- q^{m})^{2}}\,. 
\end{equation}
This formula has a simple interpretation in terms of the gauge theory in Section~\ref{sec:gaugetheory}: it deals with the massive modes of the ghost fields for the bosonic gauge symmetry and in the denominator we have the modes for the local supersymmetry.  
In a similar way it is possible to derive a localization formula for the character of the chiral structure sheaf~(\ref{eq:eulerssh}). One simply omits all the fermionic modes, which is done by setting $y=0$ in~(\ref{eq:zero}). 

Separating the zero-mode contribution\footnote{The use of contour integrals in order to express the zero mode contribution in fixed point theorem has been recently adopted in the literature (see \cite{Martelli:2006yb,Butti:2006au,Forcella:2007wk}).} from the massive modes, one gets the simplified expression 
\begin{eqnarray}
\label{eq:zeroA}
  Z^{0}_{X}(y,t|q) = \frac{1}{(1-y)^{r}} \left( \prod_{a=1}^{r} \oint 
\frac{dz_{a}}{z_{a}} \right) \prod_{i=1}^{d} 
\frac{ \left(1- y\, t^{v_i}\, z^{Q_i} \right) 
 }{\left(1- t^{v_i}\, z^{Q_i} \right) 
 }
\end{eqnarray}
where the prefactor $1/(1-y)^{r}$ is needed to cancel the zero 
mode of the local supersymmetry. Depending on $Q_i$, 
we integrate around the poles 
\begin{equation}
\label{eq:poles}
1- t^{v_i}\, z^{Q_i} =0
\end{equation}
with either the positive or negative components of $Q_i$. 
For example, if there is only one gauge symmetry, i.e. $r=1$, then 
we collect the poles with non-zero charges $Q_{i}$ into the two sets 
\begin{eqnarray*}
P_{+}&=&\{{\rm set~of~solutions~of~(\ref{eq:poles})}| i\in I_{+}\}\,,\\
P_{-}&=&\{{\rm set~of~solutions~of~(\ref{eq:poles})}| i\in I_{-}\}\,. 
\end{eqnarray*}
where we defined the index sets $I_{\pm}$ in~(\ref{eq:Iplusminus}). Then, we take the residues of the integrand ${\cal I}(y,t,z)$ in (\ref{eq:zeroA}) only with respect to one set either $P_+$ or $P_{-}$. We have noticed that if there are two sets of poles we have that 
\begin{equation}
\sum_{z_{*} \in P_{+}\cup P_{-}} {\rm Res}_{z=z_{*}}({\cal I}(y,t,z))=0\,.
\end{equation} 
In other words, we have implicitly compactified the complex $z$--plane to a $\mP^1$. In the case that either $P_{+}$ or $P_{-}$ is empty, then all poles 
have to be taken into consideration. For the general case $r > 1$, the 
procedure should be iterated for each single variable $z_{a}$ with $a=1,\dots,r$. We have checked in examples that the decompostion of the set of poles into $P_{\pm}$ is independent of the choice of basis in which the charges $Q_{a}$ are defined. 

The next step is to compute the integral in (\ref{eq:zero}) which contains 
the complete contributions of the massive modes. For that integral, we 
still take the same poles as for the zero mode part.  There is still an infinite number of poles depending on $q$. It is possible to show that by choosing a single pole with $q$ the zero mode partition function is not reproduced and the pole of the zero mode part of the integrand should necessarily be taken. 
However, we can restrict the integration domain in the annulus defined by the two poles closest to the poles given by (\ref{eq:poles}) 
\begin{equation}
1- t^{v_i}\, z^{Q_i}  q =0\,, \hspace{2cm}
1- t^{v_i}\, z^{Q_i}  q^{-1} =0\,,
 \end{equation}
for $|q| > 1$ since the other poles with $q^k$ are automatically excluded from the annulus. One could have choosen the integration domain centered around the pole $1- t^{v_i}\, z^{Q_i}  q =0$, but in that case excludes the poles of the zero mode part. This gives a wrong answer since it does not reproduce the zero 
mode contribution to the partition function. 
 
%%%%%%%%%%%%%%%%%%%%%%%%%%%%%%%%%%%%%%%%%%%%%%
\subsection{The formula by Borisov and Libgober}
\label{sec:bor}
 
We briefly collect here some results of \cite{Borisov:1999ab,Borisov:2000fg} in order to compare with our result. They find a formula for the elliptic genus using the gluing procedure described in Section~\ref{sec:toric}. Starting from the covering of $X_\Sigma$ 
%the free $\beta\gamma bc$ system 
constructed from affine patches $U_{\sigma}$, $\sigma\in\Sigma$, they calculate the \v Cech cohomology of the bundle ${\cal ELL}(X_{\Sigma})$ in~(\ref{ellcomp}). The elliptic genus $\Ell_{X_\Sigma}$ is then the generating function for the dimensions of $H^0(U_\sigma,{\cal ELL}(X_{\Sigma}))$, with $U_\sigma$ running over the \v Cech complex. The result is a concise formula in terms of the defining data of the toric variety $X_{\Sigma}$:
 \begin{equation}
  \label{eq:borA}
  \Ell_{X_\Sigma}(y,q) = y^{-n/2} \sum_{m\in M} \sum_{\sigma \in \Sigma} (-1)^{{\rm codim}\, \sigma} \left( \prod_{i=1}^{\dim \sigma} \frac{1}{ 1 - y q^{v_{i}\cdot m} }\right) G(y|q)^n\,,   
\end{equation}
where the sum over $\sigma$ runs over all the cones in the fan $\Sigma$. $M$ is again the dual lattice, $v_{i}$ are the generators of the cone $\sigma$ of dimension $\dim \sigma$. The function $G(y|q)$ is defined as 
 \begin{equation}
   \label{eq:borB}
    G(y|q) = \prod_{k\geq 1} \frac{(1 - y q^{k-1}) (1 - y^{-1} q^{k}) }{ (1- q^{k})^{2}}\,.
\end{equation}
The factor $1/(1 - y q^{v_{i} \cdot m})$ cannot be expanded around $q=0$, 
but only after multiplying it by $G(y|q)$. However, in order to regularize the above expressions one can introduce a set of parameters $t$ which are precisely the coordinates in the torus $\mT$ of each single cone $\sigma$. Again, introducing these parameters means working in equivariant cohomology. The fixed points of the torus action are in 1-1 correspondence with the cones of maximal dimension, and these are the only ones 
that contribute in~(\ref{eq:borB}) after the regularization. Performing the contour integral in~(\ref{eq:zeroA}) also yields a sum over the poles that correspond to the fixed points \footnote{This statement would need more qualifications in case the variety is not Calabi-Yau}, and so our formula is consistent with the results of Borisov-Libgober. An explicit verification in some examples will be given in section ~\ref{sec:compact}.

%%%%%%%%%%%%%%%%%%%%%%%%%%%%%%%%%%%%%%%%%%

\section{Non--compact examples}
\label{sec:non-compact}

%%%%%%%%%%%%%%%%%%%%%%%%%%%%%%%%%%%%%%%%%%

\subsection{The orbifolds}
\label{sec:orbifolds}

The orbifold $\mC^2/\mZ_2$ is a singular non-compact Calabi-Yau manifold 
and we use this example to perform cross-checks of the localization formula and to understand the structure of the physical spectrum. In the first section, we 
will consider the singular space $\mC^2/\mZ_2$ and we compute the partition function 
by projecting out the contributions of the non-invariant operators. In this way, we can easily 
write the partition function and describe the corresponding vertex operators. In the 
second section, we consider the blow-up of the singular space, namely the total 
space of the bundle ${\cal O}(-2) \rightarrow \mathbb{P}^1$, and we compute the partition function 
using the localization technique and we describe the physical states in terms 
of the \v Cech cohomology. Finally, we view the same space as an 
hypersurface in $\mathbb{C}^3$ and we compute the physical vertex operators in terms 
of the \v Cech cohomology and of the BRST symmetry.  

%%%%%%%%%%%%%%%%%%%%%%%%%%%%%%%%%%%%%%%%%%

\subsubsection{The singular space $\mC^2/\mZ_2$}
\label{sec:C2suZ2}

The orbifold as the singular space 
$\mC^{2}/ \mZ_2$, is written in terms of variables $u,v$. The discrete symmetry $\mathbb{Z}_2$ acts as follows 
\begin{equation}
(u,v) \rightarrow (-u,-v)\,.
\end{equation}
The invariant combinations $x = u^{2}, y = v^{2}, z = uv$.
satisfy the constraint $xy =z^{2}$.  The action
\begin{equation}
S = \int d^2z (w_u \bar\p u + w_v \bar \p v)\,,
\end{equation}
is invariant under the discrete symmetry if the latter is extended on 
the conjugate momenta $w_u, w_v$:
\begin{equation}
  \label{eq:newC}
  w_{u} \rightarrow - w_{u}\,, \qquad
  w_{v} \rightarrow - w_{v}\,.
\end{equation}
The action is invariant under rigid symmetry generated by  
the $\mathbb{Z}_{2}$ invariant currents
\begin{align}
  \label{eq:newD}
  J_{1} &= :w_{u} u:\,, &
  J_{2} &= :w_{v} v:\,, &
  J_{3} &= w_{u} v\,, &
  J_{4} &= w_{v} u\,. &
\end{align}
Using these currents it is rather easy to compute the affine algebra $\widehat{\mathrm{sl}}(2,\mathbb{C})_{k=-1} \times \widehat{\mathrm{gl}}(1,\mathbb{C})_{k=-2}$. In terms of the variables $u,v$, we compute the untwisted partition function $Z_{\mC^{2} / \mZ_{2}}(t_{u},t_{v}|q)$ where $t_{u}$ and $t_{v}$ correspond to scaling the fields $u,v$ and the conjugates $w_{u},w_{v}$ (the symmetry is generated by $J_{1,0}=\oint dz J_{1}$ and  $J_{2,0}= \oint dz J_{2}$). If we set
\begin{equation}
  \label{eq:parA}
Z_{+} = \frac{1}{(1-t_{u})(1-t_{v})} \prod_{n=1}^{\infty} \frac{1}{
(1-t_{u} q^{n}) (1-t_{u}^{-1}q^{n})(1-t_{v} q^{n})(1-t_{v}^{-1}q^{n})}\,,  
%=  - \frac{1}{t} q^{1/6} \frac{\eta(\tau)^{2}}{\theta_{1}(z,\tau)^{2}}   
\end{equation}
we have
\begin{eqnarray}
  \label{eq:parB}
  Z_{\mC^{2}/\mZ_{2}} &= &\frac{1}{ 2}\left(Z_{+} (t_{u},t_{v}|q) + Z_{+}(-t_{u},-t_{v}|q)\right) \\
  &=& \frac{1+ t_{u} t_{v} }{ (1- t_{u}^{2})(1- t^{2}_{v})} + q \frac{(t_{u}+ t_{v})^{2}  (1+ t_{u} t_{v})}{ t_{u} t_{v} (1 - t_{u}^{2})(1- t^{2}_{v})}+ O(q^{2})\cr
  &=&\frac{1+t^{2} }{ (1-t^{2})^{2}} + q \frac{4 (1+t^{2}) }{ (1-t^{2})^{2}} + q^{2}\frac{1+t^{2} }{ (1-t^{2})^{2}} \Big(\frac{3}{ t^{2}} + 8 + 3 t^{2}\Big) + O(q^{3})\nonumber
\end{eqnarray}
where we have set $t_{u}=t_{v}=t$ in the last line. 
It is interesting to identify each single operator of the vertex algebra to see 
how the counting works. At the zero mode level we have all 
combinations of $u^{p}$ and $v^{q}$ 
that are invariant under the $\mathbb{Z}_{2}$ symmetry. 
At the first massive level we have the operators 
\begin{eqnarray}
(\p u, \p v) \rightarrow 2 t\,, ~~~~~~~~~~~~~~~~
(w_{u}, w_{v}) \rightarrow 2/t\,, 
\end{eqnarray}
multiplied by the zero modes $u^{p} v^{q}$. Since the new operators 
appear linearly, and they transform under the $\mathbb{Z}_{2}$ symmetry 
we need to multiply them by the non-invariant sector 
of the zero modes $2t/(1-t^{2})^{2}$. 
At the second massive level we have the new vertex operators
\begin{eqnarray}\label{secondlevel}
&&(w_{u} \p u, w_{u} \p v, w_{v} \p u, w_{v} \p v) \times (\p u, \p v) \rightarrow 4 \nonumber \\
&&((\p u)^{2}, (\p v)^{2}, \p u \p v) \rightarrow 3 t^{2}\,, \nonumber\\
&&(w^{2}_{u}, w_{v} w_{u}, w^{2}_{v}) \rightarrow 3/t^{2}\,, \\
&&(\p^{2} u, \p^{2} v) \rightarrow 2 t\,, \nonumber\\
&&(\p w_{u}, \p w_{v})  \rightarrow 2/t\,, \nonumber
\end{eqnarray}
and they have to be multiplied by the zero modes. The first 
three sets on new operators, being $\mathbb{Z}_{2}$ invariant, have 
to be paired with invariant zero modes $(1 + t^{2})/(1-t^{2})^{2}$, while 
the last two lines have to be multiplied by the non-invariant zero 
modes $2 t/(1- t^{2})^{2}$. In the forthcoming section, we identify 
these operators with those in other descriptions of the orbifold. 

%%%%%%%%%%%%%%%%%%%%%%%%%%%%%%%%%%%%%%%%%%

\subsubsection{The resolved orbifold ${\cal O}(-2) \rightarrow \mP^{1}$}
\label{sec:res}

The resolved orbifold is identified by the toric diagram construct with the 
vectors of the lattice $\mZ^2$
\begin{align}
{n}_{1} &= (1,1)\,, &
{n}_{2} &= (-1,1)\,, &
{n}_{3} &= (0,1)\,. 
\end{align}
The corresponding $(P|Q)$ matrix (\ref{eq:PQ}) is easily constructed by noting the relation ${n}_{1}+{n}_{2}-2 {n}_{3}=0$ between these vectors, which implies that the charges of the gauge symmetry are $(1,1,-2)$. The total charge $\sum_{i} Q_{i} =0$ since it is a Calabi-Yau manifold. 

The orbifold described in this way is seen as a toric variety $X=\left(\mC^{3} \setminus F\right) / \mC^{*}$ and we can use the global homogeneous coordinates $a_{i}$ with $i=1,2,3$ with the $\mC^*$-action $(1,1,-2)$: 
\begin{align}
a_{1} &\rightarrow \Lambda a_{1}\,, &
a_{2} &\rightarrow \Lambda a_{2}\,, &
a_{3} &\rightarrow \Lambda^{-2} a_{3}\,.
\end{align}
where $\Lambda$ is the local gauge parameter and $F = \{a_{1}=a_{2}=0\}$. We also add the conjugate momenta $p_{i}$ which scale with the charges $(-1,-1,2)$. In order that the free action $S= \int d^{2}z \sum_{i=1}^{3} p_{i} \bar\p a_{i}$ is invariant under the gauge symmetry, the fields must solve the constraint 
\begin{equation}
  \label{eq:parC}
  :p_{1} a_{1}: + :p_{2} a_{2}: - 2 :p_{3} a_{3}: =0\,. 
\end{equation}
In terms of $a_{i}$ coordinates, we construct the $\mT$-invariants monomials (known as characters and denoted by $\chi^m$ in Section~\ref{sec:toric}) 
\begin{align}
\label{mapA}
x &= a_{1}^{2} a_{3}\,, &
y &= a_{2}^{2} a_{3}\,, & 
z &= a_{1}a_{2}a_{3}\,,
\end{align}
which satisfy the constraint $xy =z^{2}$. In addition, using the map (\ref{mapA}) we can easily compute the pull-back on the conjugate momenta  $w_{x}, w_{y}, w_{z}$
\begin{align}
p_{1} &= 2 w_{x} a_{1} a_{3} + w_{z} a_{2} a_{3}\,, &
p_{2} &= 2 w_{y} a_{2} a_{3} + w_{z} a_{1} a_{3}\,, &
p_{3} &= w_{x} a^{2}_{1} + w_{y} a^{2}_{2}+ w_{z} a_{1} a_{2}\,.
\end{align}
to find the relation with the momenta. In the same way as above, the action is invariant under the independent rescaling of the three fields $ a_{i} $. 
To compare with~(\ref{eq:parB}) we use the assignments given by the vectors namely $t_{1}$ for the first column of the $(P|Q)$ matrix, $t_{2}$ for the second column and $z$ for the gauge symmetry and we changed the variables as follows
\begin{align}
t_{1} \sqrt{t_{2}} &= t_{u}\,, &
\sqrt{t_{2}}/t_{1}  &= t_{v}\,, &
z \sqrt{t_{2}} &= \Lambda_{0}\,.
\end{align}

Let us compute the partition function using the localization formula~(\ref{eq:zero}). Notice that the computation in~(\ref{eq:parB}) can be viewed as a localization on the two fixed points of the orbifold action, but here we would like to implement the full $\mC^*$ symmetry. Following~(\ref{eq:parB}) we have
\begin{eqnarray}
  \label{eq:parD}
  Z_{X}(t_{u},t_{v}|q) &=& \oint \frac{d\Lambda_0}{ \Lambda_0} \frac{1}{ (1 - t_{u} \Lambda_0)(1 - t_{v} \Lambda_0)(1 -  \Lambda_0^{-2})} \times\\
  &&\prod_{n\geq 1} \frac{(1- q^{n})^{2} }{ (1 - t_{u} \Lambda_0 q^{n})  (1 - \frac{1}{ t_{u} \Lambda_0} q^{n}) (1 - t_{v} \Lambda_0 q^{n})(1 - \frac{1}{ t_{v} \Lambda_0} q^{n})(1 - \Lambda_0^{-2} q^{n}) (1 - \Lambda_0^{2} q^{n})}\cr
  &=&\frac{1}{ (1 - t_{u})(1 - t_{v})} \prod_{n\geq 1} \frac{1 }{ (1 - t_{u} q^{n})(1 - \frac{1}{ t_{u}} q^{n})(1 - t_{v} q^{n})(1 - \frac{1}{ t_{v}} q^{n})} \cr
  &&+  (t_{u} \rightarrow - t_{u}, ~t_{v} \rightarrow - t_{v}) \,.
\end{eqnarray}
which coincides with~(\ref{eq:parB}). Notice that we have taken only the two poles $\Lambda_0= \pm 1$ associated to the vectors $n_1$, $n_2$ with positive $Q_i$ as in the prescription in Section~\ref{sec:localization}. The numerator $\prod_{k\geq 1}(1-q^n)^{2}$ represents the contribution of ghost fields. 

We can repeat the computation by adding the fermions (we denote by 
$  \Ell_{X}(t_{u},t_{v},y|q)$ the quantities evaluated in the context of the chiral-de Rham theory). 
We skip the details and we provide the final result
\begin{equation}
  \label{eq:finA}
  \Ell_{X}(t_{u},t_{v},y|q) = y \left( \frac 
{\theta_{1}(y \, t_{u} |q) \theta_{1}(y\, t_{v} |q) }{ \theta_{1}( t_{u} |q) 
\theta_{1}( t_{v} |q) } + \frac{\theta_{1}(- y \, t_{u} |q) \theta_{1}(- y \, t_{v} |q) }{ \theta_{1}(-t_{u} |q) \theta_{1}(- t_{v} |q)  }\right)\,.  
\end{equation}
whose zero modes contribution is 
\begin{equation}
\label{ellA}
 \Ell^{0}_{X}(t_{u},t_{v},y) =
 \frac{1 + t_u\,t_v - {\left( t_u + t_v \right) }^2\,y + 
    t_u\,t_v\,\left( 1 + t_u\,t_v \right) \,y^2}{\left( 1 - {t_u}^2 \right) \,
    \left( 1 - {t_v}^2 \right) }
\end{equation}
which vanishes for $y=1$ since it is supersymmetric.\footnote{The partition 
function in (\ref{ellA}) has a nice interpretation as a two dimensional quantum field theory living in $\mathbb{C}^{2}/\mathbb{Z}_{2}$. Let us take the 
point with $t_{u}=t_{v}=\sqrt{t}$ and multiplying the new expression by 
$(1-t^{2})^{2}$, it yields $(1+t) ( 1 - 4 \,t \,y/(1+t) + t \,y^{2} )$. The first 
term is just a single scalar, the last term is the 2d auxiliary field for a 
2d chiral superfield. The term $4 t y/(1+t)$ denotes a Majorana 2d field. 
So we have a chiral superfield and the denominator $(1-t)^{2}$ are 
just $2d$ derivatives on the superfield.} (We recall that 
the point $y=1$ is the supersymmetric point since there the contribution 
of the bosonic and fermionic modes cancel). 

To compare the above result with a direct computation of the cohomology, 
we proceed to define the resolved orbifold in terms of a covering \cite{Witten:2005px}.\footnote{A similar analysis has been performed for $\mP^{1}/\mZ_{2}$ in~\cite{Tan:2006by}.} 

The space can be covered with two patches $U_{1}$ and $U_{2}$. 
In $U_{1} = \{a_{1} \neq 0\}$,  
\begin{align*}
\gamma^0 = \frac{a_{2}}{a_{1}} \,, ~~~~~~
\gamma^1 = a_{1}^{2} a_{3}\,, ~~~~ 
\end{align*}
In  $U_{2} = \{a_{2} \neq 0\}$,  
\begin{align*}
\tilde\gamma^0 = \frac{a_{1}}{a_{2}} \,, ~~~~~~
\tilde \gamma^1 = a_{2}^{2} a_{3}\,, ~~~~ 
\end{align*}
On the intersection $U_{1}\cap U_{2}$, the variables are related 
by $\tilde \gamma^0 = 1/\gamma^0$ and $\tilde \gamma^1 = \left(\gamma^0\right)^{2} \gamma^1$. 

We have a pair of systems $(\beta_i,\gamma^i)$, $i=0,1$ for the base and fibre respectively. We will be a little more general and assume that $X$ is the line bundle ${\cal O}(-k)$ with $k>0$. This is Calabi-Yau only for $k=2$.  
Then the transition functions are 
\begin{align}
\tilde \gamma^0 = & \frac{1 }{ \gamma^0}  &
\tilde \gamma^1 = & \left(\gamma^0\right)^{k} \gamma^1 \,.
\end{align}
We can easily find the classical part of $\tilde \beta_0, \tilde \beta_1$, and let's assume the general form 
\begin{align}
\label{eq:momenta}
\tilde \beta_0 = & - \beta_0 \left(\gamma^0\right)^{2} + k \gamma^0 \gamma^1 \beta_1 + A(\gamma^1,\gamma^0) \partial \gamma^0 + B(\gamma^1,\gamma^0) \partial \gamma^1 \,, \\
\tilde \beta_1 = & \frac{1 }{ \left(\gamma^0\right)^{k}} \beta_1 + C(\gamma^1,\gamma^0) \partial\gamma^0 + D(\gamma^1,\gamma^0) \partial \gamma^1 \,.\nonumber
\end{align}
The requirement of the absence of singularities in the OPEs yields the following conditions: 
\begin{align} 
A = & 2 + \frac{k^{2}}{ 2} + k \frac{\gamma^1 }{ \gamma^0} B \, , &
C = & \frac{1}{ \left(\gamma^0\right)^{k+2}} B \, &
D = & 0 
\end{align} 
and $B$ is left undetermined, which should correspond to a symmetry in the system -- in fact, the indeterminacy is of the form 
\begin{align}
\tilde \beta_0 \rightarrow & \tilde \beta_0 + \frac{B }{ \left(\gamma^0\right)^{k}} \partial \tilde \gamma^1 \,,  &
\tilde \beta_1 \rightarrow & \tilde \beta_1 - \frac{B }{ \left(\gamma^0\right)^{k}} \partial \tilde\gamma^0\,.
\end{align}
The symmetry is generated by the holomorphic 2-form $F=B \left(\tilde \gamma^0\right)^{k} d\tilde\gamma^0 \wedge d\tilde \gamma^1$. This is well-defined on both patches only if $k=2$, when it becomes the holomorphic form of the CY. Then assuming $B=0$ we get 
\begin{align}
\label{eq:beta}
\tilde \beta_0 = & - :\beta_0 \left(\gamma^0\right)^{2}: + k \gamma^0 :\gamma^1 \beta_1: + (2 + \frac{k^{2} }{ 2}) \partial \gamma^0 \,, &
\tilde \beta_1 = & \frac{1 }{ \left(\gamma^0\right)^{k}} \beta_1 \,.
\end{align}
Let us consider  operators that are globally defined sections of the sheaf. At dimension zero, they are generated by
\begin{align}
\left(\tilde \gamma^0\right)^{a} \left(\tilde \gamma^1\right)^{b}, \, & \quad b \geq 0 \,, 0 \leq a \leq k b \, \,; \nonumber \\
Z_{l=0} = & \frac{1+(k-1) t}{(1-t)^{2}}
\end{align}
For $k=2$, we get  agreement with (\ref{eq:parB}). For instance one can check the 3 operators with scaling 1: $x=\left(\tilde\gamma^0\right)^2 \tilde \gamma^1, \, y = \tilde \gamma^1, \, z = \tilde \gamma^0 \tilde \gamma^1$. At dimension 1, (we list the operators and their contribution to the character for $k=2$)
\begin{align}
\left(\tilde\gamma^0\right)^{a} \left(\tilde \gamma^1\right)^{b} \partial \tilde\gamma^0 \,, \quad & b \geq 0 \,, 0 \leq a \leq kb-2 \,, &
& \frac{t+ t^{2}}{(1-t)^{2}} \nonumber \\
\left(\tilde\gamma^0\right)^{c} \left(\tilde \gamma^1\right)^{d} \partial \tilde \gamma^1 \,, \quad & d \geq 0 \,, 0 \leq c \leq k(d+1)-1 \,, &
& \frac{2 t}{(1-t)^{2}} \nonumber \\
\left(\tilde \gamma^1  \left(\tilde\gamma^0\right)^{k}\right)^{d} \partial \left(\tilde \gamma^1 \left(\tilde \gamma^0\right)^{k}\right) \,, \quad & d \geq 0 \,, &
& \frac{t-t^{2}}{(1-t)^{2}} \nonumber \\
\left(\tilde \gamma^0\right)^{a} \left(\tilde \gamma^1\right)^{b} \tilde \beta_1 \,, \quad & b \geq 0, \, 0 \leq a \leq k(b-1) \,,  &
& \frac{1+t}{(1-t)^{2}} \nonumber \\
\left(\tilde \gamma^0\right)^{c} \left(\tilde \gamma^1\right)^{d} \tilde \beta_0 \,, \quad & d \geq 0, \, 0 \leq c \leq kd \,,  &
& \frac{1+t}{(1-t)^{2}} \nonumber \\
\left(\tilde \gamma^0\right)^{kd -1} \left(\tilde \gamma^1\right)^{d} \left(\left(\tilde \gamma^0\right)^{2} \tilde \beta_0 + (2+\frac{k^{2} }{ 2}) \partial \tilde \gamma^0\right) \,, \quad & d \geq 0 \,,  &
& \frac{(1-t)}{(1-t)^{2}} \nonumber \\
\left(\gamma^1\right)^{d} \beta_0 \,, \quad & d \geq 0 \,,  &
& \frac{(1-t)}{(1-t)^{2}} \nonumber 
\end{align}
Summing all contributions, 
\begin{equation}
Z_{l=1} = \frac{4 (1+t)}{(1-t)^{2}} 
\end{equation}
the same as  (\ref{eq:parB}). This is in agreement with the general statement that the elliptic genus of an orbifold coincides with that of a crepant resolution (that is, a resolution that preserves the canonical bundle. In present example, this amounts to preserving the CY condition)~\cite{Borisov:2000fg}.  

%%%%%%%%%%%%%%%%%%%%%%%%%%%%%%%%%%%%%%%%%%

\subsubsection{$xy - z^{2}=0$ in $\mathbb{C}^{3}$}
\label{sec:hyper}

We consider the chiral theory described by the fields $x,y,z$ and their conjugates $w_{x}, w_{y}, w_{z}$, the action is 
\begin{equation}
  \label{eq:actA}
  S =\int d^{2}z \Big(w_{x} \bar \p x + w_{y} \bar \p y + w_{z} \bar \p z \Big)\,,
\end{equation}
which is defined up to the constraint $xy = z^{2}$ and it is invariant under 
the gauge symmetry 
\begin{align}
  \label{eq:actB}
  \delta w_{x} &= \eta y\,, & \delta w_{y} &= \eta x\,, &\delta w_{z} &= -2 \eta z\,, 
\end{align}
where $\eta$ is a commuting gauge parameter. This is also true for the energy-momentum tensor $T = w_{x} \p x + w_{y}  \p y + w_{z} \p z$ and for the gauge invariant operators 
\begin{align}
  \label{eq:actC}
  J_{1} &= w_{x} x + \frac{1}{ 2} w_{z} z\,, &
  J_{2} &= w_{y} y + \frac{1}{ 2} w_{z} z\,, &
  J_{3} &= w_{x} z + \frac{1}{ 2} w_{z} y\,, &
  J_{4} &= w_{y} z + \frac{1}{ 2} w_{z} x\,,  
\end{align}
It is obvious that the four currents form an affine algebra for the non-compact  group $\rm{SL}(2) \times \mC^*$. Notice that the action, the energy-momentum tensor, the currents and the constraint $xy = z^{2}$ are holomorphic in the coordinates $x,y,z$. In addition, the constraint is homogeneous. 

It is instructive to study the cohomology in the present setting. We obtain this by translating the operators obtained in Section (\ref{sec:C2suZ2}) into the new variables. In terms of the coordinates $x=u^2, y=v^2, z=uv$ in the action~(\ref{eq:actA}), we have that
\begin{equation}
  \label{eq:newB}
S =\int d^{2}z \Big( (2 w_{x} u + w_{z} v) \bar\p u + 
(2 w_{y} v + w_{z} u) \bar\p v \Big)\,, 
\end{equation}
and therefore it yields 
\begin{align*}
  w_{u} &\equiv (2 w_{x} u + w_{z} v)\,, &
  w_{v} &\equiv (2 w_{y} v + w_{z} u)\,, 
\end{align*}
which are gauge invariant under~(\ref{eq:actB}) using the identifications of the old coordinates. We have used the gauge symmetry~(\ref{eq:actB}) 
to get rid of one of the conjugates $w_{x},w_{y}$ and $w_{z}$. 

The counting of zero modes is very simple and it can be done following~\cite{Grassi:2005jz} by observing that the possible monomials are of the form $x^{m} y^{n}$ or of the form $z x^{m} y^{n}$ we can count them obtaining 
$1+t^{2}/(1-t^{2})^{2}$. 
Let us consider the first massive operators 
\begin{eqnarray}
\label{maA}
&&(w_{u} u, w_{v} v, w_{u} v, w_{v} u)\longrightarrow J_{i}\,, ~i=1,\dots,4 \nonumber 
\\
&&(u\p u, u\p v, v \p u, v\p v) \longrightarrow 
\left(\p x, \p y, \p z, \frac{z}{2} \p \ln\left(\frac{x}{y}\right)\right)
\end{eqnarray}
multiplied by the zero modes. The last vertex operator 
$\omega \equiv {z}/{2} \p \ln({x}/{y})$ appears in the 
translation from $u,v$ variables to $x,y,z$ variable and it seems singular. 
However, we will show that it is well defined globally on the space. 
Notice that if we had used the coordinates $a_{i}$ of the previous section we would have got the form 
$\omega = (a_{1} a_{3} \p a_{2} - a_{2} a_{3} \p a_{1})$ which is a smooth holomorphic one-form.  
It is interesting to see that computing the external differential of $\omega$ we get 
\begin{equation}
\label{domega}
\Omega = d \omega = - \frac{dx \wedge dy}{z}\,,
\end{equation}
which is the holomorphic two-form of the Calabi-Yau space.\footnote{To 
compute this form we use a simple technique~\cite{Berkovits:2004px}: 
denote $x_{\mu} = (x,y,z)$. The hypersurface $xy - z^{2}=0$ can be 
written as $x_{\mu} g^{\mu \nu} x_{\nu}=0$ where the matrix $g^{\mu\nu}$ 
is symmetric and has only the non-vanishing entries 
$g^{01} =g^{10}=1, g^{22} = -1$. We define the measure by saying
\begin{equation}\label{domegaB}
dx_{\mu} \wedge dx_{\nu} = [{\cal D}x] F_{[\mu\nu]}(x)\,,
\end{equation}
 where $F_{[\mu\nu]}(x)$ is an antisymmetric tensor which satisfies 
 $x_{\rho} g^{\rho \mu} F_{[\mu\nu]}(x)=0$. This has a solution (modulo 
 reparametrizations) 
 $F_{[\mu\nu]}(x) = \epsilon_{\mu\nu\sigma} g^{\sigma\tau} x_{\tau}$ and 
 finally we have 
 \begin{equation}
 [{\cal D}x]  = \frac{\bar x^{\rho} 
 (g^{-1})_{\rho\mu} \epsilon^{\mu\nu\sigma} dx_{\mu} \wedge dx_{nu}}{\bar x^{\mu} x_{\mu}} 
\end{equation}
where $\bar x^{\mu}$ is an auxiliary vector and $ [{\cal D}x]  $ is 
independent of it. By choosing $\bar x^{\mu} = (0,0,1)$,  $ [{\cal D}x]  $ 
reduces to (\ref{domega}). 
} The Calabi-Yau form is nowhere vanishing and 
it is globally defined away from the singularity. 

In order to study if the operators in (\ref{maA}) form the complete 
set of operators at the massive level, we decompose the space into two patches:
\begin{align}
U_{1} &= \{x\neq 0, y = z^{2}/x\}\,, &
U_{2} &=\{y\neq 0, x= z^{2}/y\}
\end{align}
(in order to decompose the singular space into patches the 
singularity at $x=y=z=0$ is removed). On these patches the 
form $\omega$ becomes 
\begin{align}
\omega_{(1)} &= z \frac{\p x}{x} - \p z\,, &
\omega_{(2)} &= - z \frac{\p y}{y} + \p z\,, 
\end{align}
and therefore it is globally defined. In order to see 
if we can construct further globally defined forms 
we write the most general forms in the two patches
\begin{align}
\omega_{(1)} &= f_{1}(x,z) \p x + g_{1}(x,z) \p z\,, &
\omega_{(2)} &= f_{2}(y,z) \p y + g_{2}(y,z) \p z\,, 
\end{align}
and we get the equations for the coefficients
\begin{align}
f_{1}(x,z) &=  - \frac{z^{2}}{x^{2}} f_{2}\left( \frac{z^{2}}{x},z \right)\,, &
g_{1}(x,z) &= - \frac{ 2 \, z}{x} 
f_{2}\left( \frac{z^{2}}{x},z \right) + g_{2}\left( \frac{z^{2}}{x},z \right)\,.
\end{align}
The functions $f_{1}, g_{1}$ can be singular in $x$, but not in $z$. Vice versa 
$f_{2}$ and $g_{2}$ have to be defined in $z$ and they can be singular in $y$. 
It easy to see that there is only one non-polynomial solution which 
is given by $\omega$. The presence of the operator $\omega$ in the 
spectrum is rather unexpected since it is not of a conventional form 
for vertex operators and it does not belong to the cohomology computed on 
polynomials of the fields $x,y,z$ and their conjugates. 

We move on to the second level. Here, we have the set of operators 
(\ref{secondlevel}). Some of them can be easily expressed 
in terms of the currents $J_{i}$ and in terms of derivative 
of the first massive operators (\ref{maA}). However, 
there are new quantities
\begin{eqnarray}\label{1paA}
(w^{2}_{u}, w_{v}^{2}, w_{u}w_{v}) &\longrightarrow& 
\left(:4 w_{x}^{2} x + w_{z}^{2} y + 4 w_{x} w_{z} z:\,, \dots \right) \nonumber \\
((\p u)^{2}, (\p v)^{2}, \p u \p v) &\longrightarrow& 
\left(\tfrac{1}{4 x} (\p x)^{2}\,, 
\tfrac{1}{4 y} (\p y)^{2}\,, \tfrac{1}{4 z} (\p x) (\p y)\right)\,. 
\end{eqnarray}
The first operators have negative powers of the scaling parameters. 
They indeed contribute to the partition functions (\ref{eq:parB}) 
the terms $3/t^{2}$. More importantly, these vertex operators are not 
constructed in terms of the currents $J_{i}$. 
The second set of operators are two-forms 
and they have to be understood again in term of the patches. 
Let us take the first example $(\p u)^{2}_{(1)}$ on the first patch and we 
use the mapping to go to second patch to get 
\begin{equation}\label{2pa}
(\p u)^{2}_{(2)} = \frac{1}{4 y} (\p z)^{2} + \frac{z^{2}}{4 y^{3}} (\p y)^{2} - 
\frac{z}{2 y} \p z \p y\,.
\end{equation}
which is smooth and therefore it is globally defined. In the same way as above, 
we can study whether these are the only globally defined vertex operators.\footnote{We have referred several times to the analogy 
of the present models with the ghost sector of pure spinor string theory. 
We would like to point out here that the complete analysis of gauge 
invariant operators is not yet performed and it would be very interesting 
to see whether operators of the form (\ref{maA}) and (\ref{1paA}) appear also in that 
framework. The invariance under the Poincar\'e symmetry might 
constrain the system in such a way that there are no problems as 
works for the anomalies \cite{Nekrasov:2005wg}.} 

%%%%%%%%%%%%%%%%%%%%%%%%%%%%%%%%%%%%%%%%%%

\subsubsection{The orbifold $\mC^2/\mZ_3$}
\label{sec:C2_Z3}

It is useful to consider another example of orbifold where the toric description is obtained by introducing a double $\mC^*$ action. Let us consider the general case for the moment of $\mC^{2}/\mZ_{n}$ described by the action of ${\mZ}_{n}$ on ${\mC}^{2}$ as 
\begin{equation}
  \label{eq:orA}
  (u, v) \rightarrow (e \, u, e^{n-1}\, v)\,, ~~~~ e = e^{2 \pi i/n}  
\end{equation}
whose generators are $n_{1} = (n-1,1)$ and $n_{2}= (-1,1)$ which are related by the conditions $n_{1} + (n-1) n_{2}=0\, ~{\rm mod} \, n$ for each component. 

We can compute the partition function by using the above construction, namely 
by computing the partition functions for free fields $u,v$ and then applying the orbifold projection. However, since it is clear that this method works, we use 
this example to discuss multiple actions of ${\mC}^{*}$. For that we 
follow the construction in~\cite{Lust:2006zh} and we derive the $n-1$ new vectors as follows
\begin{equation}
  \label{eq:orB}
  w_{i} = \frac{i}{ n} n_{1} + \frac{n-i}{ n} n_{2} = (i - 1,1)\,, \qquad i = 1, \dots, n-1\,.  
\end{equation}
The blown-up geometry is $X_{\Sigma} = \left(\mC^{n+1} \setminus F_{\Sigma} \right)/\left(\mC^{*}\right)^{n-1}$ (the case discussed above is ${\mC}^{2}/{\mZ_{2}}$ which is the blow-up to $\left(\mC^{3} \setminus F_{\Sigma} \right)/\mC^{*}$ with a single $\mC^{*}$ action). 
  
The general formula is not very illuminating, here we discuss the example ${\mC}^{2}/\mZ_{3}$ in order to explain our prescription of choosing the poles in the case of a gauge group with several factors. The elliptic genus can be computed to be
%\begin{eqnarray}
 % \label{eq:orC}
 % Z^{\rm cdr}_{\mC^{2}/\mZ_{3}} &=& y^{2} G^{-2}(y|q) 
  %\int \frac{d\Lambda_{0} }{ \Lambda_{0}} 
  %\frac{d\Gamma_{0} }{ \Gamma_{0}} 
  %\times\cr
 % & &
 % \frac{\theta_{1}(y\, \Lambda_{0} \Gamma_{0}^{2} t_{u}^{2} t_{v} |q)
 %\theta_{1}(y\, \Lambda_{0}^{2} \Gamma_{0} t_{u}^{-1} t_{v}|q)
 %\theta_{1}(y\, \Lambda^{-3}_{0} t_{v}|q)
 %\theta_{1}(y\, \Gamma^{-3}_{0} t_{u} t_{v}|q)
  %}{ 
 %\theta_{1}( \Lambda_{0} \Gamma_{0}^{2} t_{u}^{2} t_{v} |q)
 %\theta_{1}( \Lambda_{0}^{2} \Gamma_{0} t_{u}^{-1} t_{v}|q)
 %\theta_{1}( \Lambda^{-3}_{0} t_{v}|q)
 %\theta_{1}( \Gamma_{0}^{-3} t_{u} t_{v}|q)
  %}\,,
%\end{eqnarray}
%where 
%\begin{equation}
 % \label{eq:gigg}
 % G(y|q) = \prod_{k\geq 1} \frac{(1- y q^{k-1})(1- y^{-1} q^{k}) }{ (1- q^{k})^{2}}\,.
%\end{equation}
%In order to make formula~(\ref{eq:orC}) more transparent, we use a change of coordinates and we get 
\begin{eqnarray}
  \label{eq:orD}
  \Ell_{\mC^{2}/\mZ_{3}}(t_1,t_2,y|q) && = y^{2} G(y|q)^{-2} 
  \int \frac{d\Lambda_{0}}{ \Lambda_{0}} 
  \frac{d\Gamma_{0} }{ \Gamma_{0}} 
  \times\cr
  && \frac{\theta_{1}(y\, \Lambda_{0}  t_{1} |q)
 \theta_{1}(y\, \Gamma_{0} t_{2}|q)
 \theta_{1}(y\, \Lambda_{0} / (\Gamma_{0})^{2}|q)
 \theta_{1}(y\, \Gamma_{0} / (\Lambda_{0})^{2}|q)
  }{ 
 \theta_{1}(\Lambda_{0}  t_{1} |q)
 \theta_{1}(\Gamma_{0} t_{2}|q)
 \theta_{1}(\Lambda_{0} / (\Gamma_{0})^{2}|q)
 \theta_{1}(\Gamma_{0} / (\Lambda_{0})^{2}|q)
   } 
\end{eqnarray}
where $\Gamma_{0}$ and $\Lambda_{0}$ are the zero modes of the two gauge symmetries. To compute the integral (\ref{eq:orD}) we can proceed as follows: one first performs the integral over $\Gamma_0$ and for that one can choose either the positive or the negative poles. The after the computation of the residues is done, the resulting integrand depends on the new poles and then, one can choose the most convenient set of residues. It can be checked that the choices of the set of the poles as well as of the order of integration all lead to the same result. 

%%%%%%%%%%%%%%%%%%%%%%%%%%%%%%%%%%%%%%%%%%

\subsection{The bundles $\cO_{\mP^1}(n)$}
\label{sec:O_n}

The next simplest toric spaces that can be studied are the bundles ${\cal O}_{\mP^1}(n)$. For those we construct the chiral algebra starting from free theory on a single patch and by gluing the vertex operators on the patches intersections. The simplest example with $n=-2$ has already appeared in the previous section as the resolution of the singularity $\mC^2/\mZ_2$. 

We again denote by $\gamma^0$ and $\beta_0$ the coordinate on $\mP^1$ and its momentum and with $\gamma^1$ and $\beta^1$ the coordinate and the momentum on the fiber. We define the currents (see~(\ref{eq:beta}) with $n=-k$)
\begin{align}
  \label{eq:Oncurrents}
  J_+ &= \betat_0\,, & J_- &= \beta_0\,, & J_3 &=-\beta_0\gamma^0-\frac{n}{2}\beta_1\gamma^1\,, & K_m &= (\gamma^0)^m \beta_1\,. 
\end{align}
The currents $K_m$ behaves under change of coordinate patch as
\begin{equation}
  \label{eq:reflection}
  \widetilde{K}_m = K_{n-m} \,.
\end{equation}
They satisfy the following algebra
\begin{align}
  \label{eq:Onalgebra}
  J_+(z)J_+(w) &= \textrm{reg}\,, & J_-(z)J_-(w) &= \textrm{reg}\,, \notag\\
  J_+(z)J_-(w) &= -\frac{2+\frac{n^2}{2}}{(z-w)^2} + \frac{2J_3(w)}{z-w}\,, & K_{m_1}(z)K_{m_2}(w) &= \textrm{reg}\,,\notag\\
  J_+(z)K_m(w) &= (m-n)\frac{K_{m+1}(w)}{z-w}\,, & J_-(z)K_{m}(w) &= -m\frac{K_{m-1}(w)}{z-w}\,,\notag\\
  J_3(z)K_{m}(w) &= \left(m-\tfrac{n}{2}\right) \frac{K_m(w)}{z-w}\,, & J_3(z)J_3(w) &= -\frac{2}{(z-w)^2}\,, \notag\\
  J_3(z)J_+(w) &= \frac{J_+(w)}{z-w}\,, & J_3(z)J_-(w) &= -\frac{J_-}{z-w}\,.
\end{align}
We have to distinguish two cases: $n\geq 0$ and $n < 0$. For $n\geq 0$ we see that the occuring values for $m$ are $m=0,\dots,n$. Furthermore, from~(\ref{eq:reflection}) we see that the $n+1$ currents $K_m$ are globally defined. Therefore they define sections in $H^0(\mP^1, \css{\mP^1}(n))$ and are in one-to-one correspondence with their zero modes in $H^0(\mP^1, \cO_{\mP^1}(n))$. For $n <0$, on the other hand, the currents $K_m$ are not globally defined, hence there are no sections. 

We immediately recognize the $\widehat{\ts\tl(2)}$ subalgebra generated by $J_\pm, J_3$ which was already present for $\mP^1$~\cite{Witten:2005px}. For $n\geq 0$ the underlying Lie algebra of zero modes of~(\ref{eq:Onalgebra}) is the semidirect product of $\widehat{\ts\tl(2)}$ and its representation of spin $n$. Note that a similar extension of the algebra of $\widehat{\ts\tl(2)}$ has already been studied in~\cite{Feigin:1990ut}, cf. (2.10) there. The currents $J_\pm, J_3$ correspond to the currents $E,F,H$ there if we identify the fields $a(z),\,a^*(z),\,b(z)$ with $\beta_0(z),\,\betat_0(z),\,\beta_1\gamma^1(z)$, respectively, and set $\nu=n, \chi=0$.

Note that the currents in~(\ref{eq:Oncurrents}) can be multiplied by arbitrary powers of $\gamma^1$. The inclusion of these additional currents will yield a much bigger algebra which however does not contain further information about the geometry. It would interesting to understand whether this algebra is related to a $W$ algebra.

Next, we count the number of operators with fixed mass level. Here, we assume $n\geq 0$. The unique massless current is 1. At level 1, we have the $n+4$ currents in~(\ref{eq:Oncurrents}) with $J_3$ charge $\frac{n}{2},\frac{n-1}{2},\dots,-\frac{n}{2}$ and $+1$, $0$, $-1$. At level 2 they satisfy the non--trivial relations:
\begin{align}
  \label{eq:Oplus2relations}
  J_+K_{m-1} - J_{-}K_{m+1} = 2J_3K_m\,, \qquad\qquad m =1,\dots,n-1\,.
\end{align}
(If we include currents obtained by multiplying by powers of $\gamma^1$, there are further relations at this level.) Therefore we have the following currents at level 2:
\begin{gather}
  \left(J_+\right)^2, \left(J_-\right)^2, \left(J_3\right)^2, K_{m_1}K_{m_2}, \notag\\
  J_+K_m, J_3J_+, J_+J_-, J_3J_-, J_-K_m \notag\\
  \partial J_+, \partial J_3, \partial J_-, \partial K_m.
\end{gather}
and in addition, for $n=0,1$, $J_3K_m$.
%
%{\bf In principle, one could write down the first few terms in the expansion. In fact, if (3.46) is the only relation, we should be able to write down the generating function for the Poincar\'e polynomial.}

% It turns out that the partition function is
% \begin{align}
%   \label{eq:Oplusngenus}
%   Z_{\cO(n)}(t|q) &= 
% \end{align}

The bundles ${\cal O}(n) \rightarrow \mP^{1}$ are also interesting for the question which poles have to be taken into account. They are not Calabi-Yau spaces (only in the case $n=-2$ the first Chern class vanishes). Their total spaces are toric varieties characterized by the vectors 
\begin{eqnarray}
{n}_{1} &=(1,-n)\,, \\
{n}_{2} &=(-1,0)\,, \\
{n}_{3} &=(0,1)\,, 
\end{eqnarray}
satisfying the relation ${n}_{1} + {n}_{2} +n {n}_{3}=0$ with the charges $(1,1,n)$. These spaces are non-singular and noncompact. 

The corresponding partition function for the zero modes is given by 
\begin{eqnarray}
Z^{0}_{{\cal O}(-n)} = 
\oint \frac{d\Lambda_{0}}{\Lambda_{0}} \frac{1}{(1 - t_{1} t_{2} \Lambda_{0})
(1 - \Lambda_{0} /t_{1} ) (1 - t_{2}/ \Lambda^{n}_{0})}\,.
\end{eqnarray}
The integrand has two types of poles: the positive poles 
$\Lambda_{0} = t_{1}$ and $\Lambda_{0} = 1/t_{1} t_{2}$ and 
the negative ones $\Lambda^{n}_{0} = t_{2}$. The number of negative 
poles depends upon $n$, but it can be shown that the result is opposite 
to the result with the positive poles. 
For exampe of ${\cal O}(-3) \rightarrow \mP^{1}$, we have the 
poles $\{t_{1}, 1/t_{1} t_{2}\}$ and $\{ t_{2})^{1/3},
(t_{2})^{1/3} e^{2 \pi i /3}, t_{2})^{1/3} e^{4 \pi i /3}\}$ and the contribution 
of the zero modes is 
\begin{equation}\label{omeno3}
Z^{0}_{{\cal O}(-3)} = 
\frac{1+ q_{1} q_{2}(q_{1}+q_{2})}{(1- q_{1}^{3})(1- q_{2}^{3})}
\end{equation}
where the change of variables $t_{1} \rightarrow q_{2}^{1/11}/q_{1}^{10/11}\,, 
t_{2} \rightarrow q_{2}^{3/11} q_{1}^{3/11}$ have been used. Notice 
that the 
result describes an infinite number of vertex operators by expanding (\ref{omeno3}) and this is due to fact the space is non-compact. 
It is straightforward also to compute the elliptic genus given by 
the integral 
\begin{eqnarray}
&&\Ell_{{\cO}(-n)}(y, t_{1}, t_{2}|q) = 
G^{-1}(y|q) \oint \frac{d\Lambda_{0}}{\Lambda_{0}} 
\left[\frac{(1 - y t_{1} t_{2} \Lambda_{0})
(1 - y \Lambda_{0} /t_{1} ) (1 - y t_{2}/ \Lambda^{n}_{0})}
{(1 - t_{1} t_{2} \Lambda_{0})(1 - \Lambda_{0} /t_{1} ) (1 - t_{2}/ \Lambda^{n}_{0})} \times \right.\cr
&&
\hspace{-1.4cm}\left.\prod_{k=1}^{\infty} 
 \frac{(1 - y t_{1} t_{2} \Lambda_{0} q^{k})
(1 - y \Lambda_{0} /t_{1}  q^{k}) (1 - y t_{2}/ \Lambda^{n}_{0}  q^{k}) 
(1 - (y t_{1} t_{2}\Lambda_{0})^{-1}  q^{k})
(1 - (y \Lambda_{0} /t_{1})^{-1}  q^{k}) 
(1 - (y t_{2}/ \Lambda^{n}_{0})^{-1}  q^{k})}
{(1 - t_{1} t_{2} \Lambda_{0}  q^{k})
(1 - \Lambda_{0} /t_{1}  q^{k}) 
(1 - t_{2}/ \Lambda^{n}_{0}  q^{k})
(1 - (t_{1} t_{2} \Lambda_{0})^{-1}  q^{k})
(1 - (\Lambda_{0} /t_{1})^{-1}  q^{k}) 
(1 - (t_{2}/ \Lambda^{n}_{0})^{-1}  q^{k})} \right]
\,. \nonumber
\end{eqnarray}
 The factor $G(y|q)$ implements the gauge symmetry and it 
 removes the unwanted operators. It would be nice 
 to find an interpretation for the higher order contribution of the power 
 series in $q$ from quantum field theory.

%%%%%%%%%%%%%%%%%%%%%%%%%%%%%%%%%%%%%%%%%%

\subsection{The conifold}
\label{sec:conifold}

Another interesting model is the conifold. It admits two toric descriptions, which are dual to each other~\cite{Kreuzer:2006ax}: the first is from the point of view of the lattice $M$. Take coordinates $x,y,z,t$ of ${\mC}^{4}$ and impose the constraint
\begin{equation}
  \label{eq:coniA}
  x y - t z =0\,.  
\end{equation}
We also introduce the conjugate momenta $w_{x}. w_{y}, w_{z}$ and $w_{t}$ with the gauge symmetry
\begin{align}
  \label{eq:coniB}
  \delta w_{x} &= y\,, & \delta w_{y} &= x\,, &\delta w_{z} &= -t\,, & \delta w_{t} &= - z\,, 
\end{align}
The action is again a free action which is gauge invariant and the equations 
of motion  imply that all fields are holomorphic. The action is invariant under the rigid symmetry generated by the rescaling of the fields. The second way is from the point of view of the lattice $N$. We view the conifold as the singular toric variety $X_0= {\mC}^{4} / \mC^{*}$ with homogeneous coordinates $w_{i}$ ($i=1,2,3,4$) (and their conjugate momenta $p_{i}$) with the gauge invariances 
\begin{align}
  \label{eq:coniC}
  w_{1} &\to \Lambda w_{1}\,, &
  w_{2} &\to \Lambda w_{2}\,, & 
  w_{3} &\to \Lambda^{-1} w_{3}\,, &
  w_{4} &\to \Lambda^{-1} w_{4}\,.
\end{align}
The conjugate momenta $p_{i}$ have to satisfy the constraint $\sum_{i=1}^{4} r_{i} (p_{i} w_{i})=0$ with $r_{i}=(1,1,-1,-1)$. One choice for the vectors in the lattice $N$ is 
\begin{align}
{n}_{1} &= (0,0,1)\,, &
{n}_{2} &= (1,1,1)\,, &
{n}_{3} &= (0,1,1)\,, &
{n}_{4} &= (1,0,1)\,, 
\end{align}
and they satisfy the relation ${n}_{1} + {n}_{2} - {n}_{3} - {n}_{4}=0$ and therefore the associated charge vector is $Q=(1,1,-1,-1)$. The space is Calabi-Yau and it is singular. The singularity can be resolved by blow-ing it up in two different ways since the fan $\Sigma$ admits two different triangulations. The resulting two maximal cones correspond to the two fixed points of the torus action (see \cite{Nekrasov:2004vw}). After the blow-up the toric variety can be described as $X =\left({\mC}^4\setminus F\right)/\mC^*$ where $F= \{w_1 = w_2=0\}$ for one resolution and $F= \{w_3 = w_4=0\}$ for the other. The localization computation done with the two different triangulations leads to the same result (this has been noticed also in \cite{Martelli:2006yb}; here we extend their conclusions for each massive mode). For the elliptic genus this is a manifestation of the fact that the elliptic genus is invariant under crepant resolutions~\cite{Borisov:2000fg}.

We assign the coordinates $t_{1}, t_{2}, t_{3}$ to the action of the torus $\mT$ and we find that the partition function is 
\begin{eqnarray}
  \label{eq:coniD}
   Z_{X}(t_{1}, t_{2}, t_{3}|q) &=& \oint \frac{d\Lambda_0 }{ \Lambda_0} \frac{1}{ (1 - t_{1} \Lambda_0)(1 - t_{2} \Lambda_0)(1 - t_{3} \Lambda_0^{-1})(1 - \Lambda_0^{-1})} \times\cr
  &&  \prod_{n \geq 1} \frac{(1- q^{n})^{2}  }{  (1 - t_{1} \Lambda_0 q^{n})(1 - t_{2} \Lambda_0 q^{n}) (1 - t_{3} \Lambda_0^{-1} q^{n})(1 -  \Lambda_0^{-1} q^{n})} \times\cr
  && \frac{1}{  (1 - \frac{1}{ t_{1} \Lambda_0}q^{n})(1 - \frac{1}{ t_{2} \Lambda_0} q^{n})(1 - t^{-1}_{3} \Lambda_0 q^{n})(1 - \Lambda_0 q^{n})} 
\end{eqnarray}
We compute the integral by taking into account the two poles associated to the positive charges, $\Lambda_0=1$ and $\Lambda_0=t_{3}$, and the residues give
\begin{eqnarray}
  \label{eq:coniE}
&=& 
  \prod_{n\geq 1} \frac{1}{ (1-t_{3} q^{n}) (1-t^{-1}_{3} q^{n})} \times \cr
&&  
  \left( \frac{t_{3} }{ (t_{3} -1)} 
  \frac{1}{ (1 - t_{1})(1 - t_{2})} \prod_{n\geq1} 
  \frac{1}{ (1 - t_{1} q^{n})(1 - t_{2} q^{n})
  (1 - t^{-1}_{1} q^{n})(1 - t^{-1}_{2} q^{n})
  } + \right.\\
&& 
  \left.\frac{1 }{ (1-t_{3})} 
  \frac{1}{ (1 - t_{1} t_{3})(1 - t_{2} t_{3})} \prod_{n\geq1} 
  \frac{1}{ (1 - t_{1} t_{3} q^{n})(1 - t_{2} t_{3} q^{n})
  (1 - \frac{1}{ t_{1} t_{3}} q^{n})(1 - \frac{1}{ t_{2} t_{3}} q^{n})
  } \right) \nonumber
\end{eqnarray}
and using the relation (\ref{eq:acD}) it yields 
\begin{equation}
  \label{eq:coniF}
   Z_{X} (t_{1},t_{2},t_{3}|q)= -i \sqrt{\frac{t_{3}}{ t_{1} t_{2}}} q^{-\frac{1}{4}} 
   \frac{\eta^{3}(q)}{ \theta_{1}(t_{3}|q)}  
   \left( \frac{1}{ t^{2}_{3}\theta_{1}(t_{1} t_{3}|q) \theta_{2}(t_{2}t_{3}|q)} - \frac{1}{\theta_{1}(t_{1}|q) \theta_{2}(t_{2}|q)}\right)
\end{equation}
One interesting limit is $t_{1}\rightarrow t, t_{2} \rightarrow t$ and $t_{3} \rightarrow 1$. Notice that each term is singular in that limit, but 
$Z_{X} (t_{1},t_{2},t_{3}|q)$ has a finite limit.  The first term of the $q$-expansion contains the information about zero modes and it 
leads to the known result~\cite{Nekrasov:2004vw}:
\begin{equation}
Z^{0}_{X} (t_{1},t_{2},t_{3}) = 
\frac{(1- t_{1} t_{2} t_{3})}{(1-t_{1})(1-t_{2})(1-t_{1} t_{3})(1-t_{2} t_{3})}\,.
 \end{equation}
The computation for the chiral de Rham complex can be done analogously and the zero mode part yields in the above limit 
\begin{equation}
\Ell^{0}_{X} (t,t,1) = 
\frac{\left( 1 - t \right) \,\left(1 - t\,y \right) \,\left( 1 + t + y + \left( -6 + t \right) \,t\,y + t\,\left( 1 + t \right) \,y^2 \right)}{(1-t)^{4}}\,.
 \end{equation}
In the limit $y\rightarrow1$ it leads to the finite result $+2$. This means that the model is not supersymmetric, however by first taking the limit $t \rightarrow 1$ one can see the restoration of the supersymmetry since the polynomial becomes $(1 - 3 y + 3 y^{2} - y^{3})$ times the zero mode contribution $(1 - t^{2})/(1-t)^{4}$.  

As in the orbifold case, we can study the cohomology of the space 
in a different way. This amounts to choose a set of patches of the space. Since 
the space is singular we have to remove the tip of the cone and 
then we can cover the space with two patches $x\neq 0$ and 
$z\neq 0$. Following the analysis performed in Section~\ref{sec:hyper} one finds several forms with singular behaviour at the tip of the cone as for instance:
\begin{equation}\label{formA}
\omega = 
\frac{z}{x} dx\wedge dt - \frac{t}{x} dx\wedge dz
\end{equation}
whose differential gives the top form of the space $\Omega = d \omega = 
\frac{dx\wedge dt\wedge dz}{x}$. 
In the same we can compute 1-forms: 
\begin{equation}
\omega' = \ln(x) d(z\, t),,  
\end{equation}
whose differential is $\Omega' = \frac{z}{x} dx\wedge dt + 
\frac{t}{x} dx\wedge dz$. 

%%%%%%%%%%%%%%%%%%%%%%%%%%%%%%%%%%%%%%%%%%

\section{Compact examples}
\label{sec:compact}

%%%%%%%%%%%%%%%%%%%%%%%%%%%%%%%%%%%%%%%%%%

\subsection{$\mP^1$}
\label{sec:P1andP2}

Further interesting examples are compact toric varieties. The prototype is ${\mP}^{1}$ and has been studied in~\cite{Witten:2005px}. In the present section we consider the chiral algebra constructed on that space and compute the partition function and the elliptic genus.
 
The action is again 
\begin{equation}
  \label{eq:cpA}
   S = \int d^{2}z \beta \bar\p \gamma\,.  
\end{equation}
$\gamma$ is the coordinate on north-pole patch $U_{+}$ of the $\mP^1$. It is related to the coordinate $\gammat$ on the south-pole patch $U_{-}$ by $\gammat = 1/\gamma$. The conjugate momenta $\betat$ are related to the  $\beta$ by a $\gamma$-dependent transformation: $\betat = - \gamma^{2} \beta + 2 \p \gamma$. The second term is needed to guaratee that $\betat(z) \betat(w) \sim 0$. The energy momentum tensor is simply given by $T = \beta \p \gamma$ on the patch $U_{+}$. On a single patch the theory is free. However, going from one patch to another one has to check which quantities are globally defined. 
 
As a toric variety $\mP^1$ can be written in the form $\left(\mC^2\setminus F\right)/ {\mC^*}$ with $F=\{x_0=x_1=0\}$. The defining vectors are $n_1= 1$ and $n_2=-1$ and therefore the partition function can be computed as in the non-compact cases. The symmetry ${\mC^*}$ acts on the homogeneous coordinates by $(x_{0}, x_{1}) \rightarrow (\Lambda x_{0}, \Lambda x_{1})$. 
Thus, we can compute 
\begin{eqnarray}
  \label{eq:cpB}
   Z_{\mP^1}(t|q) &=& \oint \frac{d \Lambda_{0} }{ \Lambda_{0}} 
 \frac{1}{ (1 - t \Lambda_{0})(1 - \frac{\Lambda_{0}}{ t})}
 \prod_{k\geq1}  \frac{(1 - q^{k})^{2} }{ 
 (1 - t \Lambda_{0} q^{k}) 
 (1 - \frac{1 }{ t \Lambda_{0}} q^{k}) 
 (1 - {\Lambda_{0}}{ t} q^{k}) 
 (1 - {t }{ \Lambda_{0}} q^{k}) 
 }\cr
 &=&\frac{1}{ (1- t^{2})} \prod_{k\geq 1} {1}{ (1- t^{2} q^{k})(1- \frac{1}{ t^{2}} q^{k})} + \frac{1}{ (1- \frac{1}{ t^{2}})} \prod_{k\geq 1} \frac{1}{ (1- t^{2} q^{k})(1- \frac{1}{ t^{2}} q^{k})}\cr
 &=&  \prod_{k\geq 1} \frac{1}{ (1- t^{2} q^{k})(1- \frac{1}{ t^{2}} q^{k})}
\end{eqnarray}
where the integral is computed as the residue at the two poles $z = t$ and $z = 1/t$. Notice that the numerator takes into account the gauge degrees of freedom coming from the gauge symmetry $\Lambda$. The zero mode of the gauge transformation is removed by integrating over $\Lambda_{0}$. The result coincides with the character computed in sec. 5.8 of~\cite{Malikov:1998dw}
\begin{eqnarray}
  \label{eq:cpC}
  Z_{\mP^1}(1|q) = 
  \mathrm{Ch}(\mP^1; \css{\mP^1}) = \sum_{n \geq 0} \chi(\mP^1; \css{\mP^1})_{(n)}) q^{n}\,,
\end{eqnarray}
where $\mathrm{Ch}(X; \css{X})$ is the Euler character of $\css{X}$ and  $\css{X}$ is the sheaf of chiral vertex operators on $X$. $\chi(X; \css{X})_{(n)}$ counts the number of states at the $n$-th level. 

In the case $X = {\mP}^{1}$, there are only two cohomology groups:  $H^{0}(\mP^1, \css{\mP^1})$ and $H^{1}(\mP^1, \css{\mP^1})$. We have $H^{0}(\mP^1, \css{\mP^1})^{*} \simeq H^{1}(\mP^1, \css{\mP^1})$ and it can be checked that the isomorphism is obtained by multiplying $H^{0}(\mP^1, \css{\mP^1})$ by $\theta \equiv \p\gamma/\gamma$. This means that 
\begin{equation}
  \label{eq:cpE}
   \mathrm{Ch}(H^{0}(\mP^1,\css{\mP^1}) = \frac{1}{1-q}  \mathrm{Ch}(\mP^1; \css{\mP^1})  = \frac{1}{1-q} \prod_{k\geq 1} \frac{1}{ (1- t^{2} q^{k})(1- \frac{1}{ t^{2}} q^{k})}\,,  
\end{equation}
which concides with what Malikov has found in the study of the reducible Verma module of affine $\mathrm{sl}(2)$ algebra at the critical value $k = -2$~\cite{Malikov:1989ab}. The computation is based on the work of Feigin and Frenkel~\cite{Feigin:1990ut}.
 
Before considering a new space, we discuss the supersymmetric case. We add the supersymmetric partners of $\gamma$ and we have to impose the local gauge (super)symmetry to have a single boson and  a single fermion. The best way to do it is to observe that the equivalence $(x_{0}, x_{1}) \sim  (\Lambda x_{0}, \Lambda x_{1})$ can be extended to the  fermions $(\psi_{0}, \psi_{1})$ (see section 
\ref{sec:supersymmetric}). They are the supersymmetric variation of the  bosonic variables and we add also the supersymmetry variation  of the gauge parameter $\Lambda$. This means that we have to add to the partition function the contribution coming from the fermions (weighted with $t$ and $1/t$ for the gauge symmetry and $y$ for the fermion number). So, the partition function for the zero modes is 
\begin{equation}
  \label{eq:cpfA}
   Z^{0}_{{\mP}^{1}}(t,y) = \oint \frac{d \Lambda_{0} }{ \Lambda_{0}} 
 \frac{(1 - t y \Lambda_{0})(1 - {y \Lambda_{0} / t}) }{ (1-y) (1 - t \Lambda_{0}) (1- {\Lambda_{0} / t} )} = 1 + y\,.
\end{equation}
Here we have taken into account the two poles $\Lambda_{0} = t, 1/t$. The factor $1-y$ is coming from the zero modes of the ghost field for the local supersymmetry obtained by the variation of the gauge parameter $\Lambda_{0}$. Notice that the ghost field of the supersymmetry gauge parameter is homogeneous it has zero degree and it carries only the fermion degree. 
 
In the same way we can compute the complete elliptic genus by adding the non-zero modes 
\begin{eqnarray}
  \label{eq:cpfB}
 &&  \Ell_{{\mP}^{1}}(y,t|q) = \oint \frac{d \Lambda_{0} }{ \Lambda_{0}} \frac{(1 - t y \Lambda_{0})(1 - {y \Lambda_{0} / t}) }{ (1 - t \Lambda_{0}) (1- {\Lambda_{0} / t} )}  \times \\
  &&\prod_{k} 
 \frac{
 (1 - t y \Lambda_{0} \, q^{k})
 (1 - \frac{1}{ t y \Lambda_{0}} \, q^{k})
 (1 - \frac{\Lambda_{0}}{ t y} \, q^{k})
 (1 - \frac{t y }{ \Lambda_{0}} \, q^{k})
 }{ 
 (1 - t \Lambda_{0} \, q^{k})
 (1 - \frac{1}{ t \Lambda_{0}} \, q^{k})
 (1 - \frac{\Lambda_{0} }{ t} \, q^{k})
 (1 - \frac{t }{ \Lambda_{0}} \, q^{k})
 }
 \frac{1}{ 1-y}
 \prod_{l} 
 \frac{(1- q^{l})^{2}}{ (1 - y \, q^{l})(1 - 1/y \, q^{l})} \cr
 &&= \frac{(1 - y t^2)}{ (1- t^2) } 
 \prod_{k} \frac{ (1 - y t^{2} \, q^{k})(1 - y^{-1} t^{-2} \, q^{k})}{  (1- t^{2}\, q^{k})(1-t^{-2}\, q^{k})} +  \frac{(1 - y t^{-2})}{ (1- t^{-2}) } 
 \prod_{k} \frac{ (1 - y t^{-2} \, q^{k})(1 - y^{-1}  t^{2} \, q^{k})}{ 
 (1- t^{2}\, q^{k})(1- t^{-2}\, q^{k})} \nonumber
 \,.
\end{eqnarray}
Separating the non-zero modes from the zero modes, we get back to~(\ref{eq:cpfA}). The last factor of the second line is the contribution coming from the ghost fields (both the fermionic and the bosonic ones). We want to compare our result with the Borisov-Libgober formula (\ref{eq:borA}).
 In the case of ${\mP}^{1}$ it yields
\begin{eqnarray}
  \label{eq:borC}
&&  \Ell_{\mP^1}(y,t|q) = 
  \sum_{m \in {\mZ}} \left[ \frac{t^{m} + t^{-m} }{ (1- y q^{m})} \right] G(y|q) \\
  && = \frac{(1 - y t)}{ (1- t) } 
  \prod_{k} \frac{ (1 - y t \, q^{k})(1 - y^{-1} t^{-1} \, q^{k})}{ 
  (1- t \, q^{k})(1- t^{-1}\, q^{k})} + 
  \frac{(1 - y t^{-1})}{ (1- t^{-1}) } 
  \prod_{k} \frac{ (1 - y t^{-1} \, q^{k})(1 - y^{-1}  t \, q^{k})}{ 
  (1- t\, q^{k})(1- t^{-1}\, q^{k})} \,. \nonumber
\end{eqnarray}
which coincides with~(\ref{eq:cpfB}) provided that $t \rightarrow t^{2}$. 
Passing from the first to the second line we have used the identity
\begin{equation}
  \label{eq:borD}
   \prod_{i=1,\dots,d} \prod_{k\geq 1} 
 \frac{(1 - y t^{m_{i}} q^{k-1}) (1 - y^{-1} t^{- m_{i}} q^{k}) }{ (1- t^{m_{i}} q^{k-1}) 
 (1- t^{-m_{i}} q^{k})} = \sum_{m \in M} t^{m }\prod_{i=1,\dots,d} \frac{1}{ (1- y q^{m \cdot n_{i}})} \, G(y|q)^{d}
\end{equation}
where $n_{i}$ are the generators of the cone $\sigma$ of dimension $d$. This formula has been proven by Borisov and Libgober in~\cite{Borisov:1999ab}. One interesting observation is how to extract the zero mode part from the first line of~(\ref{eq:cpfB}). Notice that at each positive $m$ there is a factor in the infinite product that cancels the denominator $1 - y q^{k}$ and this gives a contribution to the zero mode part. In the first line
of the equality in (\ref{eq:borC}), the expression for the summation over $\mathbb{Z}$ is a generalization of the 
Borisov-Libgober (\ref{eq:borA}) where the equivariant parameter $t$ has been introduced. 
 
We still need to prove the equivalence in the limit $t \rightarrow 1$. For that purpose we need to take carefully the limit in~(\ref{eq:cpfB}) and we find the 
result 
\begin{equation}
  \label{eq:borPone}
  \mathrm{Ell}_{\mP^{1}}(y|q) = y^{- 1/2} (1- y^{2}) \sum_{m \in {\mZ}} \frac{1}{ \prod_{i=1}^{2}(1 - y q^{m \cdot n_{i}})} G(y|q)
\end{equation}
which coincides with formula~(\ref{eq:borA}) if we chose the fan generated by $n_{1} = 1, n_{2} = -1$. The fan $\Sigma$ contains three cones, the two generated by $n_{1}$ and $n_{2}$ and the zero dimensional one of the origin. Again the product $m\cdot n_{i}$ uses the usual definition of the pairing between the lattice $N_{\mR}$ and its dual $M_{\mR}$. 
 
In order to compare with the direct computation of the cohomology we need Poincar\'e duality~(\ref{eq:poinA}) yielding the isomorphism between $H^{0}({\mP}^{1}, \cdr{\mP^1})^{*}$ and $H^{1}({\mP}^{1}, \cdr{\mP^1})$. By computing the relation between a single operator of $H^{0}$ and one of $H^{1}$, we see that we have to multiplying an operator of $H^{0}$ by $c$. This means that for $t=1$
\begin{eqnarray}
  \label{eq:cpfC}
   \mathrm{Ch}(H^{1}({\mP}^{1};\cdr{\mP^1}))(y|q) &=& - y  \mathrm{Ch}(H^0({\mP}^{1}; \cdr{\mP^1}))(y^{-1}|q) \cr
  \Ell_{\mP^1}(y|q) &=&  \mathrm{Ch}H^0({\mP}^{1}; \cdr{\mP^1})(y|q)  + y  \mathrm{Ch}H^1({\mP}^{1}; \cdr{\mP^1})(y^{-1}|q) \cr
  &=& \prod_{k} \frac{ (1 - y \, q^{k})(1 - y^{-1} \, q^{k})}{  (1- q^{k })^2}\,.  
\end{eqnarray}
By computing the first two levels we have $\mathrm{Ch}(H^{0})(y|q) = 1 + q (2 - y - 1/y) + O(q)$. The first term coincides with the only dimension zero section globally defined on ${\mP}^{1}$ which is the constant. The second term can be better be written as $q ( 3 - y - 1 - 1/y)$ where we recognize the six currents at dimension 1 which are globally defined. Notice that three of them are bosonic and the others are fermionic. They indeed coincides with the superaffine $sl(2)$ found in~\cite{Tan:2006qt}. 
 
 \subsection{$\mP^2$}
 
Let us consider the space ${\mP}^{2}$. This is the toric variety $\left(\mC^{3}\setminus F\right)/\mC^*$ with a ${\mC}^{*}$ action with weights $(1,1,1)$ and $F=\{z_1=z_2=z_3=0\}$. In this space we have two charges $t_{i}$ with $i=1,2$ and the coordinates are weighted with the vectors $n_{1} = (1,0)$, $n_{2} = (0,1)$ and $n_{3}= (-1,-1)$. Then, we can easily compute the  zero mode partition function by a localization formula
\begin{equation}
  \label{eq:cpdA}
   Z^{0}_{\mP^2} = \oint \frac{d\Lambda_{0} }{ \Lambda_{0}} \frac{1}{ (1 - t_{1} \Lambda_{0}) (1 - t_{2} \Lambda_{0}) (1 - t^{-1}_{1}  t^{-1}_{2} \Lambda_{0})} = 1
\end{equation}
which is the correct value. However, to see a more interesting partition function, we need to add the fermions on the worldsheet and to implement the gauge symmetry then we have 
\begin{equation}
  \label{eq:cpdB}
   \Ell^{0}_{\mP^2} = \oint \frac{d\Lambda_{0} }{ \Lambda_{0} (1-y)} 
   \frac{(1 - y\, t_{1} \Lambda_{0}) (1 -  y\, t_{2} \Lambda_{0}) 
 (1 - y\, t^{-1}_{1}  t^{-1}_{2} \Lambda_{0}) 
 }{ (1 - t_{1} \Lambda_{0}) (1 - t_{2} \Lambda_{0}) (1 - t^{-1}_{1}  t^{-1}_{2} \Lambda_{0})} = 1 + y + y^{2}\,.
\end{equation}
The factor $1/(1-y)$ is the contribution of the zero modes of the supersymmetry 
 transformations. Again, this coincides with the classical computation of the Euler character of the space $\mP^2$. For $\mP^n$ the 
 structure sheaf for $n\geq 2$ is not defined since the $p_1(\mP^n)\neq0$ and the corresponding 
 partition function is not modular invariant. 
 
 Now, we are in position to compute  the full elliptic genus 
\begin{equation}
   \label{eq:cpdF}
 \Ell_{\mP^2} = - i \sqrt{y} q^{-\frac{1}{ 6}}\oint \frac{d\Lambda_{0} }{ \Lambda_{0}} 
 \frac{\theta_{1}(y\, t_{1} \Lambda_{0}|q) \theta_{1}( y\, t_{2} \Lambda_{0}|q) 
 \theta_{1}(y\, t^{-1}_{1}  t^{-1}_{2} \Lambda_{0}|q) 
 }{ 
 \theta_{1}(t_{1} \Lambda_{0}|q) 
 \theta_{1}(t_{2} \Lambda_{0}|q) 
 \theta(t^{-1}_{1}  t^{-1}_{2} \Lambda_{0}|q)}  \frac{\eta^{3}(q) }{ \theta_{1}(y|q)}\,.   
\end{equation}
The contour integral is evaluated at the poles $\Lambda_{0}= t_{i}$ for $i=1,2$ and $\Lambda_{0} = 1/t_{1} t_{2}$. That gives the contribution of the massive states to the Euler character. 
 
Let us now compute the same quantity using the Borisov-Libgober technique. We recall that the generators of the maximal cones are $n_{1} = (1,0)$, $n_{2} =(0,1)$ and $n_{3}= (-1,-1)$. Then from the formula we have 
\begin{equation}
  \label{eq:cpdC}
   Z_{\mP^2} = \sum_{m_{1},m_{2} \in {\mZ}^2} 
 \left(  \frac{t^{m_{1}}_{1} t^{m_{2}}_{2} + t^{m_{1} - m_{2}}_{1}  t^{-m_{2}}_{2} + 
 t^{-m_{1}}_{1}  t^{m_{2}-m_{1}}_{2}
 }{ 
 (1 - y q^{m_{1}})
 (1 - y q^{m_{2}})}\right)G(y|q)^{2}
\end{equation}
and using again the Borisov-Libgober identity~(\ref{eq:borD}), one can see 
that~(\ref{eq:cpdC}) coincides with~(\ref{eq:cpdF}). Notice that again only the maximal cones contribute to the partition function. 

We have to compare it again with formula~(\ref{eq:borA}) in the case of ${\mP}^{2}$ in the limit $t_1, t_2 \rightarrow 1$. With a careful treatment of the various limits, we found the result 
\begin{equation}
  \label{eq:cpdD}
  \mathrm{Ell}_{\mP^{2}}(y|q) = y^{-1} (1- y^{3}) \sum_{m \in {\mZ}^{2}} \frac{1}{ 
\prod_{i=1}^{3} (1- y q^{m \cdot n_{i}})} G(y|q)^2
\end{equation}
which is the explicit version of~(\ref{eq:borA}). 

%%%%%%%%%%%%%%%%%%%%%%%%%%%%%%%%%%%%%%%%%%

%\subsection{Hirzebruch surfaces}
%\label{sec:Hirzebruch}

%In the present section we consider another example of a compact space: 
%${\mP}_{1,1,2}$ which is compact, but is singluar. The resolution of the singularity is given by the space ${\mF}_2$. They are defined as follows: 
%$$
%n_0 = (0,1)\,, \quad
%n_1 = (1,0)\,, \quad
%n_2 = (-2,-1)\,, \quad
%(z_0,z_1,z_2) \sim (\Lambda z_0, \Lambda z_1, \Lambda^2 z_2)
%$$
%$${\mP}_{1,1,2} = \left({\mC}^3\setminus F \right) / {\mC}^*\,,\quad
%F= {z_0=z_1=z_2=0}\,.
%$$
%and 
%$$
%n_0 = (0,1)\,, \quad
%n_1 = (1,0)\,, \quad
%n_2 = (-2,-1)\,, \quad
%n_3 = (-1,0) \,, \quad
%$$
%$$
%(z_0,z_1,z_2,z_3) \sim (\Lambda z_0, \Lambda z_1, \Lambda^2 \Gamma z_2, \Gamma z_3)
%$$
%$${\mF}_{2} = \left({\mC}^4\setminus F \right) / ({\mC}^*)^2\,,\quad
%F= \{z_0=z_1=0\}\bigcup \, \{z_2=z_3=0\}\,.
%$$
%Notice that in the first case, the space is described by a single ${\mC}^*$ action 
%on the coordinates, while in the second case we need two actions. These means that 
%we have to write the localization formula with a double integral and we have also 
%to take into account the scaling properties of the gauge parameters. 

%%%%%%%%%%%%%%%%%%%%%%%%%%%%%%%%%%%%%%%%%%

\section{Non--reduced schemes}
\label{sec:Nonreducedscheme}

Let us consider a very simple model described by a single field $\gamma$ and its conjugate $\beta$ with the chiral free action
\begin{equation}
  \label{eq:simA}
  S = \int d^{2}z \beta\bar \p\gamma\,.
\end{equation}
In addition, we impose the quadratic constraint and the gauge invariance
\begin{equation}
  \label{eq:simB}
 \gamma^{2}=0\,, \qquad  \delta\beta = 2 \rho \gamma\,,
\end{equation}
Notice that we can form the gauge invariant current $J = :\beta \gamma:$ and the stress energy tensor $T = :\beta \p \gamma:$. To prove the gauge invariance of $J$, we use the constraint $\g^{2}=0$. 

Notice that from the classical point of view the theory has no propagating degrees of freedom since the constraint implies $\gamma =0$ and the gauge invariance removes $\beta$. The classical space is an algebraic variety defined as the spectrum of maximal ideals $\mathrm{Specm}\mathbb{C}[\gamma]/\langle \gamma^2=0 \rangle$  (see \cite{Fulton:1993ab}).\footnote{Note that this is a simple example of pure spinor constraints emerging in~\cite{Berkovits:2000fe} and analyzed in several dimensions in~\cite{Berkovits:2002zk}, \cite{Grassi:2005sb}, \cite{Wyllard:2005fh}, \cite{Adam:2006bt}.} 

Computing the Fock space and imposing the constraint on that space, one can see that there is a non-trivial set of states. (We use here the Fock space of states for simplicity of exposition, but the same results can be expressed in terms of vertex operators). 
The massless states for this quadric are $|0\rangle$ and $\gamma_{0}|0\rangle$ where $\gamma_{0}$ are the zero modes of the expansion $\gamma_{1} (z)= \sum_{k \in \mathbb{Z}} \g_{k} z^{k}$ with the condition that $\gamma_{-k} |0\rangle =0$ for $k > 0$. Therefore assigning the charge $+1$ to $\gamma$ and $-1$ to $\beta$, one gets zero mode partition function 
\begin{equation}
Z^{0}_{\g}(t) = (1+t)\,.
\end{equation}
This resembles the character formula for a worldsheet spinor. \footnote{
As a simple application of the present analysis, we can consider the 
quadric proposed in~\cite{Krotov:2006th}  written in terms of four $\gamma$'s as follows 
$\gamma_{1} \gamma_{2} =0\,,
\gamma_{2} \gamma_{3} =0\,,
\gamma_{3} \gamma_{4} =0\,,$
$
\gamma_{1}^{2}=0\,,
\gamma_{4}^{2}=0\,.
$
Then we can compute the character for each subquadrics with each condition $\l_{i}=0$, then we subtract the results for the case $\l_{i}=\l_{j}=0$ and so on. 
The final result is given by the function
\begin{equation}
  \label{eq:fooA}
  Z_{\gamma}^{\rm Losev}(t) = {(1+ 3t - t^{3})/1-t}\,.
\end{equation}
By adding also the fermionic zero modes $\theta_{i}$ (with $i=1,\dots,4$) in oder to have a complete physical model and the total contribution is 
\begin{equation}
  \label{eq:fooB}
  Z_\gamma^{\rm Losev}(t) = (1-t)^{4} {(1+ 3t - t^{3}) / (1-t)} 
 = 1 - 5 t^{2} + 4 t^{3} +3 t^{4} - 4 t^{5} + t^{6}\,.
\end{equation}
This coincides exactly with the table provided in~\cite{Krotov:2006th} and the counting of the states reflects their nature as commuting and anticommuting fields. The next step is to analyze the massive states to see if they have physical meaning.} 

The same technique can be used to compute the massive level. We expand the constraint  $\gamma^2 =0$ in modes 
\begin{align}
  \label{eq:modeA}
  \sum_{n\in \mathbb{Z}} \gamma_n \gamma_{-n} &=0\,, &
  \sum_{n\in \mathbb{Z}} \gamma_{n+1} \gamma_{-n}&=0\,, &
  \sum_{n\in \mathbb{Z}} \gamma_{n+2} \gamma_{-n} &=0\,, & \dots 
\end{align}
At each level, we find a new constraint. Acting on the vacuum $|0\rangle$ selects only the negative modes and we are left with 
\begin{align}
  \label{eq:modeB}
  \gamma_0 \gamma_{0} |0\rangle &=0\,, &
\gamma_{0} \gamma_{-1} |0\rangle &=0\,, &
2 \gamma_{0} \gamma_{-2} + \gamma_1^2 |0\rangle&=0\,, & \dots
\end{align}
At each level, we have a quadratic constraint to implement. Now, we consider 
the modes $\gamma_{-n}$ as independent  fields and we solve the constraints as before. 
For example, we can compute the zero mode constribution as follows: setting $\gamma_0 =0$, we have the contribution $1/(1 - t q)$ (notice that the fields $\gamma_{-1}$ carry the parameter $t$ and the conformal weight $+1$), then we select $\gamma_{-1} =0$ and we have the contribution $(1 + t)$; finally, we set$\gamma_0 = \gamma_{-1} =0$ and we get the contribution $-1$. This implies that 
the total contribution is 
\begin{equation}
  \label{eq:modeC}
  Z^{1}_{\gamma}(t) = \partial_q \left(\frac{1 + t - t^2 q }{ (1-t) (1- t q)}\right)|_{q=0}
\end{equation}
where only the coefficient of the power $q$ is taken into account. We can push this further by taking into account also the next level. However, first 
we add also the contributions due to $\beta$'s. This is done by adding the 
denominator $(1 - t^{-1} q)$ since there is no other constraint
\begin{equation}
  \label{eq:modeD}
  Z^{1}_{\gamma}(t) = \frac{1 -  t^2 - t^2 q + t^3 q - (t^{-1} -1) q }{ (1-t) (1- t q)(1- t^{-1} q)} = (1+ t) + (1+ t) q + O(q^2)\,.
\end{equation}
We can read off the constraints from the numerator: $t^2$ corresponds to the 
zero modes constraints $\gamma_0^2$. The term $t^2 q$ corresponds to the 
constraint $\gamma_0 \gamma_{-1} =0$. However, we have overcounted the constraint $\gamma_0^2 \gamma_{-1}$ which is obtained by acting with $\gamma_0$ on $\gamma_0\gamma_{-1}$ or with $\g_{-1}$ on the first constraint. The next term subtracts the gauge invariance $\delta \beta_{-1} = \Lambda_{-1} \gamma_0$ where $\Lambda_{-1}$ has conformal weight $+1$ and we have to subtract the overcounting because $\gamma_0 \delta \beta_{-1} =0$ due to the first constraint. In this way, we have a complete understanding of the formula~(\ref{eq:modeD}). The states that give the $q$ contributions are $\gamma_{-1}|0\rangle$ and $\beta_{-1} \gamma_0|0\rangle$. 

Now, we follow a different path: we solve the constraint $\gamma^{2}=0$ using a pair of fermions and we then we implement a localization formula similar to~(\ref{eq:gaC}) for their gauge invariance. 
We introduce two fermions $\xi$ and $\eta$ and set $\gamma = \xi \eta$. Of course, this factorization is defined up to the gauge symmetry:
\begin{equation}
  \label{eq:simBA}
  \xi \rightarrow \Lambda \xi\,, \qquad\eta \rightarrow \Lambda^{-1} \eta 
\end{equation}
It is obvious that the description in terms of two fermionic fields 
automatically implies that $\gamma^{n}=0$ $\forall n>1$. Using the solution 
and plugging it into the action~(\ref{eq:simA}), we have 
\begin{equation}
  \label{eq:simC}
  S = \int d^{2}z ( \beta \xi \bar\p\eta - \beta \eta \bar\p\xi ) = \int d^{2}z (p_{\xi} \bar \p\xi + p_{\eta} \bar\p \eta)   
\end{equation}
where $p_{\xi} = - \beta \eta$ and $p_{\eta} = \beta \xi$ and satisfy: $p^{2}_{\xi} = p^{2}_{\eta} = p_{\xi} p_{\eta} + p_{\eta} p_{\xi}=0$. Notice that the two conjugate momenta $p_{\xi}$ and $p_{\chi}$ are gauge invariant $\delta p_{\xi} = -\delta \beta \eta = - 2 \rho (\xi \eta) \eta =0$ and analogously for $\delta p_{\eta}$. Notice also that the OPE's of the original theory imply the 
OPE's of the fermionic model
\begin{eqnarray}
  \label{eq:simD}
  p_{\xi}(z) \xi(w) &=& - (\beta \eta)(z) \xi(w)  = - \beta(z) \left(\eta(w) + (z-w) \p \eta(w)\right) \xi(w)  \cr
  &=& - \beta(z) \left( \eta\xi(w) + (z-w) \p\eta\xi(w) \right) = - \b(z) \gamma(w) \cr
&\sim& \frac{1}{ (z-w)}\,,   
\end{eqnarray}
and in addition, we have that $p_{\xi}(z) \eta(w) \sim 0$ since $\eta^{2}=0$. Finally, we observe that 
\begin{eqnarray}
  \label{eq:simE}
   p_{\xi}(z) p_{\eta}(w) &=& -\beta(z) \eta(z) \xi(w) \beta(w) \cr 
   &=& - \beta(z) \frac{1}{ 2}\left( \eta\xi(w) + \eta(z)\xi(z) + (z-w) (\p \eta \xi(w) - \eta \p \xi)\right)\beta(w) \cr
  &\sim& - \frac{1}{ (z-w)} \b(w) + \frac{1}{ (z-w)} \b(w) = 0\,.
\end{eqnarray}
The gauge invariant current $J = :\beta \gamma: = :\beta \xi\eta: =  :p_{\eta} \eta: = :p_{\xi} \xi:$. This implies that there should be a constraint  on the conjugate momenta. Indeed, in  order that the action be gauge invariant under~(\ref{eq:simBA}), we need that
\begin{equation*}
   :p_{\xi} \xi: - :p_{\eta}\eta: =0\,.  
\end{equation*}
Let us now discuss the implication of the gauge symmetry~(\ref{eq:simBA}). We want to show that all gauge invariant combinations are indeed the operators which satisfy the constraint~(\ref{eq:simB}). For example $\gamma = \xi \eta$ and $\gamma^{2} =0$. The next level is to consider the first derivative of the constraint $\gamma \p \gamma =0$. At the first level we have the following type of vertex operators $\p\xi, \p\eta, \xi\p\xi, \eta\p\eta, \eta\p \xi, \xi\p\eta, \xi\eta\p\xi, \xi\eta\p\eta$. However, the only gauge invariant combination (as can be easily checked) is $\p(\xi\eta)$. In the same way we have
\begin{align}
  \label{eq:levA}
   l&=1\,, && \p(\xi\eta)\,,\cr
 l&=2\,, && \p^{2}(\xi\eta), \p^{2}(\xi\eta) \xi\eta\,,\cr
 l&=3\,, &&\p^{3}(\xi\eta), \p^{3}(\xi\eta) \xi\eta\,,\cr
 l&=4\,, &&\p^{4}(\xi\eta), \p^{4}(\xi\eta) \xi\eta\,, \p^{3}(\xi\eta) \p(\xi\eta)
\end{align}
The action is invariant under the separate rescaling of $\xi$ and of $\eta$ and we parametrize this rescaling by $t_{\xi}$ and $t_{\eta}$. So, we can write the partition function by imposing for the moment only zero mode part of the gauge symmetry~(\ref{eq:simBA}) (notice that the  gauge parameter $\Lambda$ is local and therefore it has a zero mode part plus  the higher modes). The zero mode part of the gauge symmetry rescales the fields as follows $\xi \rightarrow \Lambda_{0} \xi$ and $\eta \rightarrow \Lambda^{-1}_{0} \eta$. Thus, 
\begin{equation}
  \label{eq:levB}
  \hat Z_{\xi\eta}(t_{\xi}, t_{\eta}|q) = \oint \frac{d \Lambda_{0} }{ \Lambda_{0}} \prod_{n=0}^{\infty} (1- t_{\xi} \Lambda_{0} q^{n}) (1- t^{-1}_{\xi} \Lambda^{-1}_{0} q^{n+1}) (1- t_{\eta} \Lambda^{-1}_{0} q^{n})(1- t^{-1}_{\eta} \Lambda_{0} q^{n+1})
\end{equation}
Let us take only the zero mode part of $\hat Z_{\xi\eta}(t_{\xi}, t_{\eta}|q)$, which gives 
\begin{equation}
  \label{eq:levC}
   \hat Z^{0}_{\xi\eta}(t_{\xi}, t_{\eta}|q) = \oint \frac{d \Lambda_{0} }{ \Lambda_{0}} (1- t_{\xi} \Lambda_{0}) (1- t_{\eta} \Lambda^{-1}_{0}) = 1 + t_{\xi} t_{\eta} \,,
\end{equation}
which is the correct answer since it can be easily reproduced by the two vertex operators $1, (\xi\eta) =\gamma$. In order to implement the complete gauge symmetry, we need to add also the non zero mode part of the gauge symmetry and finally we have
\begin{equation}
  \label{eq:levD}
   Z_{\xi\eta}(t_{\xi}, t_{\eta}|q)  = 
 \oint \frac{d \Lambda_{0} }{ \Lambda_{0}} \prod_{n=0}^{\infty} (1- q^{n+1}) 
 (1- t_{\xi} \Lambda_{0} q^{n}) (1- t^{-1}_{\xi} \Lambda^{-1}_{0} q^{n+1})
 (1- t_{\eta} \Lambda^{-1}_{0} q^{n})(1- t^{-1}_{\eta} \Lambda_{0} q^{n+1})  
\end{equation}
leading to the expansion
\begin{equation}
  \label{eq:levE}
  Z_{\xi\eta}(t_{\xi}, t_{\eta}|q) = (1 + t_{\xi} t_{\eta}) + (1 + t_{\xi} t_{\eta}) q + (t^{-1}_{\xi} t^{-1}_{\eta} + 2 + 2 t_{\xi} t_{\eta} + t_{\xi}^{2} t_{\eta}^{2}) q^{2} + O(q^{3})
\end{equation}
which is the correct expansion to the third order. The appearance 
of negative powers of $t_{\xi}$ and $t_{\eta}$ is due to the gauge invariant 
operators $\beta^{n} \gamma$.  For a precise comparison of the above formula with the operators of the chiral algebra one has to study the complete set of operators 
written in terms of the fermions, but the best way to do it is using the bosonization formulas for a 2d system. In order to compare with the first level
computed by hand we see that we need to set $t = t_{\eta} t_{\xi}$ in the partition function (\ref{eq:modeD}).

We reported the present example, but one can generalize it to other schemes such as 
$\g^{n} =0$ where $n>2$. In that case a decomposition in terms of fermions can be done by setting $\g = \sum_{I} \xi^{I} \eta_{I}$ where $I = 1,\dots, n-1$. Again there is a gauge symmetry, but the situation is more complicated and will be discussed in a forthcoming paper on the generalization of the present work to nonabelian gauge groups.

Finally, we point out that the present example has a very interesting aspect: 
the fields are fermionic and therefore enter the numerator of the localization formula, and 
the gauge symmetry is bosonic, so it is realized with anticommuting ghosts which appear again in the 
numerator of the integrand. So, the integration over $\Lambda_{0}$ projects out the fermionic contributions leaving only the even power combintions of fields and the ghost fields expressed 
by the contribution $\prod_{n}(1-q^{n+1})$ are needed to remove the unwanted fermion contributions 
at the massive level. 

%%%%%%%%%%%%%%%%%%%%%%%%%%%%%%%%%%%%%%%%%%

\section{Supermanifolds}
\label{sec:superman}

%%%%%%%%%%%%%%%%%%%%%%%%%%%%%%%%%%%%%%%%%%

\subsection{The super-Calabi--Yau $\mathbb{P}^{(1|2)}  
\equiv \Pi({\cal O}(1)\oplus{\cal O}(1)) \rightarrow \mathbb{P}^{1}$}

We consider here the supermanifold $\mathbb{P}^{(1|2)}$. 
It is characterized by the fact that the space is a super-K\"alher, and it is super Calabi-Yau. 
This can be viewed by computing the Chern class or 
computing the super-Ricci tensor from the 
metric derived from the potential
\begin{equation}\label{CHA}
K = \ln(1 + |\gamma|^{2} + \bar\theta_{i} \theta^{i})\,,
\end{equation}

It is parametrized by one bosonic complex coordinate $\gamma$ and 
two fermionic complex coordinates $\theta_{i}$ ($i =1,2$). The bosonic 
submanifold of $\mathbb{P}^{(1|2)}$ is $\mP^{1}$ and we use the following two patches 
$$
U_{1} = \{{\gamma, \theta_{i}}\}\,, ~~~~~
U_{2} = \{{\widetilde\gamma, \widetilde\theta_{i}}\}\,, ~~~~~
$$
and on the intersection $U_{1} \cap U_{2}$, we have the identifications 
\begin{equation}\label{CHB}
\widetilde \gamma = \frac{1 }{\gamma}\,, ~~~~~
\widetilde\theta_{i} = \frac{\theta_{i} }{ \gamma}\,, 
\end{equation}
Using the fact that the action (modulo contact terms) should be equivalent 
on the two patches, we conclude that the relations between 
the conjugate momenta on the two patches are 
\begin{equation}\label{CHC}
\widetilde \b = - \gamma^{2} \b - \gamma \, p_{i} \theta^{i}\,, ~~~~
\widetilde p_{i} = \g p_{i}\,.
 \end{equation}
It is easy to show that in order to satisfy all commutation 
relations between the conjugate momenta and the fields $\widetilde \gamma$ and $\widetilde \theta_{i}$, we need to replace $\widetilde \b$ with 
\begin{equation}\label{CHD}
\widetilde \b = - \g^{2} \b - \g \, p_{i} \theta^{i} + \p \gamma\,, ~~~~
\end{equation}
Now, we can compute the chiral algebra from the sheaf of 
operators. The result is 
\begin{equation}\label{CHE}
J_{++} = - \g^{2} \b - \g \, p_{i} \theta^{i} + \p \gamma\,, ~~~
J_{3} = - 2 \gamma\beta - p_{i} \theta^{i}\,, ~~~
J_{--} = \beta\,,
\end{equation}
$$
Q_{+, i} = \gamma \, p_{i} \,, ~~~
Q_{-,i} = p_{i}\,, ~~~
K_i^j = p_i \theta^j - \delta_i^j p \cdot \theta\,,
$$
The first three generators are bosonic, while the others are 
fermionic. They have the following non-vanishing commutation 
relations 
\begin{equation}\label{CHF}
[J_{++}, Q_{+,i}] = 0\,, ~~~~~
[J_{++}, J_{3}] = - 2 J_{++}\,, ~~~~~
[J_{++}, J_{--}] =  2 J_{3}\,, ~~~~~
\end{equation}
$$
[J_{++}, Q_{-,i}] = - Q_{+,i}\,, ~~~~~
[Q_{+,i}, J_{3}] = - Q_{+,i}\,, ~~~~~
[Q_{+,i}, J_{--}] = Q_{-,i}
$$
$$
\{Q_{+,i}, Q_{-,j} \} = 0\,, ~~~~
[J_{3}, Q_{-,i}]=-Q_{-,i}\,, ~~~~
[J_{3}, J_{--}] = -2 J_{--}\,, ~~~~
[Q_{-,i}, J_{--}]=0\,.
$$
Again (as in the bosonic case) we can see that the 
algebra is not semisimple. In the present case the 
algebra is a super-Lie algebra which has $sl(2,\mR)$ as 
a bosonic subalgebra. The other two pairs 
fermionic generators $Q_{+,i}$ and $Q_{-,i}$ 
are in the fundamental representation of $sl(2, \mR)$.  
The algebra is then a semidirect product of $sl(2, \mR)$ with 
a $\Pi T^{2}$ where $T^{2}$ is a (super)-abelian ideal carrying the representation and the parity of generators of $T^{2}$ is inverted. 
Adding the generators $K_{i}^{j}$ we get that the full algebra becomes 
$\mathrm{psl}(2|2)$, the superalgebra which is the simplest generalization of the bosonic result~\cite{Gotz:2006qp}. 

There are some quadratic relations between these generators
\begin{equation}\label{CHFA}
Q_{+,i} Q_{-,j} + Q_{+,j} Q_{-} =0\,, 
\end{equation}
which can be easily verified. The level of the Kac-Moody algebra is 
$k= - 1$. 

%Let us now compute the partition functions. It 
%seems to me that the correct partition function is 
%given by
%\begin{equation}\label{CHG}
%Z(y,q) = \frac{(1 - q^{2})^{2} }{ (1-q)}  
%\prod_{k=1}^{\infty} \frac{ (1 - y q^{k})^{2} (1 - y^{-1} q^{k})^{2} 
%}{ 
%(1 - y^{2} q^{k}) (1 - y^{-2} q^{k}) 
%}
%\end{equation}

Using again the localization procedure, we can determine the partition function. The contribution of the $\theta$'s appear in the numerator since they 
are fermionic and in addition we scale them as
\begin{equation}
\theta_{1} \rightarrow y_{1} \theta_{1}\,, ~~~~~
\theta_{2} \rightarrow y_{2} \theta_{2}\,.
\end{equation}
The K\"ahler potential is invariant under this scaling since the conjugate 
variable $\bar \theta_{i}$ and $\bar \gamma$ scale in the opposite way. The gauge symmetry is the same as for $\mP^{1}$ (there is no gauge symmetry for the fermions since the model is supersymmetric only in the target space) and therefore we have
\begin{eqnarray}\label{supA}
Z_{\mP^{(1|2)}}(y_{i}, t_{1}|q) &=& \oint \frac{d\Lambda_{0}}{\Lambda_{0}} 
\frac{ (1 - y_{1} \Lambda_{0})(1 - y_{2}  \Lambda_{0})}{(1 - t_{1}  \Lambda_{0})(1 -  \Lambda_{0}/t_{1})}
\times\\
&&\prod_{k>1}
\frac{ (1 - y_{1} q^{k}\Lambda_{0})(1 - y_{2}\Lambda_{0}  q^{k} ) 
 (1 - \frac{q^{k}}{y_{1} \Lambda_{0}})(1 - \frac{q^{k}}{y_{2}\Lambda_{0}})
}{(1 - t_{1}  \Lambda_{0} q^{k})(1 -  \frac{t_{1} \, q^{k}}{\Lambda_{0}})}\nonumber\,.
\end{eqnarray}
It is rather instructive to compute explicitly the zero mode part. 
After the integration performed we get 
\begin{equation}
Z^{0}_{\mP^{(1|2)}}(y_{i}, t_{1}) = 1 - y_{1} y_{2}\,. 
\end{equation}
The interpretation of the result is interesting. The space $\mP^{(1|2)}$ has 
essentially the same cohomology as $\mP^{1}$. The contribution of the fermions 
has to be studied using the technique of superforms \cite{Grassi:2004tv}, and 
it can be argued that only the cohomology groups $H^{(0|0)}(\Omega)$ and 
$H^{(1|0)}(\Omega)$ are different from zero. The second index in $(p|q)$ denotes 
the picture number which is a second grading for the superforms. Using the 
differential operators $Y, Z$ which raise and lower the picture, it can be shown that other cohomology groups are either empty or isomorphic to the above two. It turns out that the $H^{(0|0)}(\Omega_{0})=1$ and  $H^{(1|0)}(\Omega_{0})=\theta_{1} \theta_{2}$. The latter is a 1-cochain and it can not be written as  difference between two coboundaries. In addition, it can be shown that all derivatives of $\theta_{1}\theta_{2}$ are coboundaries. 

In the more interesting case, $\mP^{(3|4)}$, we get the zero mode partition function
\begin{equation}
Z^{0}_{\mP^{(3|4)}}(y_{i}, t_{1}) = (1 - y_{1} y_{2} y_{3} y_{4})\,. 
\end{equation}
where we have added four fermions $\theta_{i}$. Notice that second term 
has the same interpretation as above. 

%%%%%%%%%%%%%%%%%%%%%%%%%%%%%%%%%%%%%%%%%%

\subsection{$\Pi({\cal O}(-1)\oplus{\cal O}(-1)) \rightarrow \mathbb{P}^{1}$}
\label{sec:super2}

The last example we consider is a again a supermanifold, but 
not Calabi-Yau. The interest for this space stems from the fact that it has non 
trivial sections and the zero mode contribution is more interesting than the superprojective 
space studied above. 

In particular, the space is formed by a basis $\mP^{1}$ 
with two patches $U_{i}$ and transforming from one patch to the other 
the fibers transform as follows 
\begin{equation}\label{caA}
\tilde \theta_{i} = \gamma \theta_{i}\,,
\end{equation}
This differs from the previous section and it allows new operators 
at each level.
 
Following the construction of the previous sections, at the massless level 
we have the following global states (they belong to the \v Cech cohomology):
\begin{equation}\label{caB}
1, (\tilde\theta_{i}, \widetilde\gamma\tilde\theta_{i}), 
(\tilde\theta_{i} \tilde\theta_{j}, \widetilde\gamma \tilde\theta_{i} \tilde\theta_{i}, \widetilde\gamma^{2} \tilde\theta_{i}\tilde\theta_{j})\,.
\end{equation}
Assigning the charge $+1$ to the fermionic fibers, we get the 
following character
\begin{equation}\label{caC}
Z^{0}_{\Pi({\cal O}(-1)\oplus{\cal O}(-1)) \rightarrow \mathbb{P}^{1}} 
= (1 -  (1+ t^2) y_1 - (1+ t^2) y_2 + (1 +  t^{2} + t^4) y_1 y_2\,.
\end{equation}
Notice that for $t =1$ the espression vanishes and this implies that 
the states are organized into a supermultiplet. In particular, there are 
four bosonic states and four fermionic states. In addition, we have to notice 
that in the previous example of 
$\Pi({\cal O}(-1)\oplus {\cal O}(-1)) \rightarrow \mP^{1}$, we have only a single 
massless operator in $H^{(0|0)}$ and therefore there is no supersymmetry in the target space. 

%%%%%%%%%%%%%%%%%%%%%%%%%%%%%%%%%%%%%%%%%%

\section{Conclusions and Outlook}
\label{sec:conclusions}

We first summarize some of the results that we have obtained in the present work. We have found a generalization of the localization formula for topological theories (i.e. restricting to the zero modes) to chiral theories, in particular including massive states  They appear in the spectrum of the chiral models based on half-twisted $(0,2)$ supersymmetric models. These models can be nicely formulated in terms of nonlinear $\beta\gamma$ (and $bc$) systems. We have explored two versions of this formula: the bosonic version for the chiral structure sheaf and the supersymmetric version for the chiral de Rham complex. In order to achieve this, we have expressed the $\beta\gamma$ system in terms of an abelian gauge theory. It turns out that toric geometry provides a natural framework. In some sense we have found a gauge theory realization of the holomorphic quotient of a toric variety. This is to be contrasted with the gauge theory realization of the symplectic quotient of a toric variety in terms of the gauged linear sigma-model~\cite{Witten:1993yc}. We checked the formula for several toric varieties, both compact and non-compact, and we compared this with the direct computation of the cohomology whenever this was possible. For other models we compare with known results always finding agreement. 

There are many open questions and further directions to be studied. Here we list a few of them.
\begin{itemize}
  \item Localization: A full derivation of our localization formula from first principles clarifying the origin of our prescription for the choice of poles is definitely highly desirable.
  \item Hypersurfaces and divisors (in toric varieties): In a future publication we plan to extend our localization formula to hypersurfaces in compact toric varieties, in particular, to Calabi-Yau hypersurfaces. For the latter, Borisov and Libgober have already found in~\cite{Borisov:1999ab} a formula for the elliptic genus of Calabi-Yau spaces in terms of the data defining the toric ambient space. We already have extended the present analysis to compute the character of divisors in non--compact toric varieties.
  \item Non-perturbative effects: These effects become important, once we go away from the large volume limit. The results of the present work only apply to this limit. Away from the large volume limit, worldsheet instantons will contribute and destroy the chiral algebra. It will be necessary to go beyond the chiral approximation and to consider the full conformal field theory. First steps in this direction have been taken in~\cite{Frenkel:2005ku}, \cite{Frenkel:2006fy}. In the first of these references, the authors also worked into realm of toric geometry.
  \item Counting BPS states in $N=1$ theories in $D=4$: recently some interesting work related to the localization formula appeared in~\cite{Martelli:2006yb,Butti:2006au,Forcella:2007wk}. Here again, non-compact toric varieties such as $Y^{p,q}$, del Pezzo surfaces, $L^{p,q,r}$ play an important role, serving as building elements for the construction of Sasaki-Einstein spaces on the gravity side. It has been argued that the partition function computed in this way reproduces the counting of a certain set of BPS states in the dual gauge theory. With the results of the present work applied to divisors in these toric varieties and taking the zero mode part, we are able to reproduce these partition functions. It would be very interesting to see whether the higher modes of the partition function of the structure sheaf (or of the chiral de Rham complex) can be interpreted as counting other operators in the dual gauge theory.
\item Koszul resolution: An open problem is the relation to the quantum Koszul resolution studied in~\cite{Grassi:2006wh}, in particular how the partition function can be obtained in that context.
\item Non-abelian gauge groups: An obvious direction is the extension of our formulism to non-abelian gauge groups. This will be the subject of a forthcoming publication. The class of target spaces can then be extended to include homogeneous spaces.
\item Pure spinors: as we have already mentioned in the introduction and several times in the main text, the present work provides a first step to compute the partition function for the full-fledged superstring in the pure spinor formalism. In precedent work \cite{Berkovits:2005hy,Grassi:2005jz} only the zero modes and the first massive levels have been studied. This is certainly the most challenging and most important problem to be studied in the future.
\end{itemize}

%%%%%%%%%%%%%%%%%%%%%%%%%%%%%%%%%%%%%%%%%%

\subsection*{Acknowledgments}
\label{sec:ack}

We are grateful to Maximilian Kreuzer, to Francisco Morales, and to Jan Troost for helpful discussions. This work is supported in part by the European Union RTN contract MRTN--CT--2004--005104 and by the Italian Ministry of University (MIUR) under the contract PRIN 2005--023102 ``Superstringhe, brane e interazioni fondamentali.''
The work of E. S. is supported by the Marie Curie Grant MERG--CT--2004--006374.

%%%%%%%%%%%%%%%%%%%%%%%%%%%%%%%%%%%%%%%%%%
\begin{small}

\end{small}


\begin{thebibliography}{99}

\bibitem{Witten:1991zz}
  E.~Witten,
  ``Mirror manifolds and topological field theory,''
  in Mirror Symmetry I, S.~T. Yau (ed.), 121--160.
  [arXiv:hep-th/9112056].
  %%CITATION = HEP-TH 9112056;%%

\bibitem{Witten:2005px}
  E.~Witten,
  ``Two-dimensional models with (0,2) supersymmetry: Perturbative aspects,''
  arXiv:hep-th/0504078.
  %%CITATION = HEP-TH 0504078;%%

\bibitem{Kapustin:2005pt}
  A.~Kapustin,
  ``Chiral de Rham complex and the half-twisted sigma-model,''
  arXiv:hep-th/0504074.
  %%CITATION = HEP-TH 0504074;%%

\bibitem{Witten:1986bf}
  E.~Witten,
  ``Elliptic genera and quantum field theory,''
  Commun.\ Math.\ Phys.\  {\bf 109} (1987) 525.
  %%CITATION = CMPHA,109,525;%%

\bibitem{Hirzebruch:1992ab}
  F.~Hirzebruch, T.~Berger, R.~Jung, 
  ``Manifolds and Modular Forms,''
  Aspects of Mathematics {\bf E 20}, Vieweg, Braunschweig, 2nd ed., (1994) 

\bibitem{Malikov:1998dw}
  F.~Malikov, V.~Schechtman and A.~Vaintrob,
  ``Chiral de Rham complex,''
  Commun.\ Math.\ Phys.\  {\bf 204} (1999) 439
  [arXiv:math.ag/9803041].
  %%CITATION = MATH-AG 9803041;%%

\bibitem{Malikov:1999ab}
  F.~Malikov, V.~Schechtman and A.~Vaintrob,
  ``Chiral de Rham complex II,''
  Differential topology, infinite-dimensional Lie algebras, and
  applications, Amer.\ Math.\ Soc.\ Transl.\ Ser.\ 2 {\bf 194} 
  (1999) 149 -- 188,
  [arXiv:math.ag/9901065].

\bibitem{Malikov:1999ac}
  F.~Malikov, and V.~Schechtman,
  ``Chiral Poincar\'e duality,''
  Math.\ Res.\ Lett. {\bf 6} (1999) 533 -- 546,
  [arXiv:math.ag/9905008].

\bibitem{Gorbounov:1999ab}
  V.~Gorbounov, F.~Malikov, and V.~Schechtman,
  ``Gerbes of chiral differential operators,''
  Math.\ Res.\ Lett. {\bf 7} (2000) 55 -- 66,
  [arXiv:math.ag/9906117].

\bibitem{Gorbounov:2000ab}
  V.~Gorbounov, F.~Malikov, and V.~Schechtman,
  ``Gerbes of chiral differential operators II,''
  Invent. Math. {\bf 155} (2004) 605--680,
  [arXiv:math.ag/0003170].

\bibitem{Gorbounov:2000ac}
  V.~Gorbounov, F.~Malikov, and V.~Schechtman,
  ``Gerbes of chiral differential operators III,''
  The orbit method in geometry and physics, Marseille, 2000,
  Progr.\ Math. {\bf 213}, (2003), 73--100,
  [arXiv:math.ag/0005201].

\bibitem{Borisov:1998dw}
  L.~A.~Borisov,
  ``Vertex algebras and mirror symmetry,''
  Commun.\ Math.\ Phys.\  {\bf 215} (2001) 517
  [arXiv:math.ag/9809094].
  %%CITATION = MATH-AG 9809094;%%

\bibitem{Borisov:1999ab}
  L.~A.~Borisov, and A.~Libgober,
  ``Elliptic genera of toric varieties and applications to mirror symmetry,''
  Invent.\ Math.\ {\bf 140}  (2000) 453--485
  [arXiv:math.AG/9904126] 

\bibitem{Borisov:2000fg}
  L.~A.~Borisov, and A.~Libgober,
  ``Elliptic genera of singular varieties,''
  Duke\ Math.\ J.\ {\bf 116}  (2003)  319--351
  [arXiv:math.ag/0007108].
  %%CITATION = MATH-AG 0007108;%%

\bibitem{Berkovits:2000fe}
  N.~Berkovits,
  ``Super-Poincare covariant quantization of the superstring,''
  JHEP {\bf 0004}, 018 (2000)
  [arXiv:hep-th/0001035].
  %%CITATION = HEP-TH 0001035;%%

\bibitem{Grassi:2001ug}
  P.~A.~Grassi, G.~Policastro, M.~Porrati and P.~Van Nieuwenhuizen,
  ``Covariant quantization of superstrings without pure spinor constraints,''
  JHEP {\bf 0210}, 054 (2002)
  [arXiv:hep-th/0112162].
  %%CITATION = HEP-TH 0112162;%%

\bibitem{Grassi:2002tz}
  P.~A.~Grassi, G.~Policastro and P.~van Nieuwenhuizen,
  ``The massless spectrum of covariant superstrings,''
  JHEP {\bf 0211}, 001 (2002)
  [arXiv:hep-th/0202123].
  %%CITATION = HEP-TH 0202123;%%
  
\bibitem{Berkovits:2005hy}
  N.~Berkovits and N.~Nekrasov,
  ``The character of pure spinors,''
  Lett.\ Math.\ Phys.\  {\bf 74}, 75 (2005)
  [arXiv:hep-th/0503075].
  %%CITATION = HEP-TH 0503075;%%

\bibitem{Grassi:2005jz}
  P.~A.~Grassi and J.~F.~Morales Morera,
  ``Partition functions of pure spinors,''
  Nucl.\ Phys.\ B {\bf 751}, 53 (2006)
  [arXiv:hep-th/0510215].
  %%CITATION = HEP-TH 0510215;%%

\bibitem{Grassi:2006wh}
  P.~A.~Grassi and G.~Policastro,
  ``Curved beta-gamma systems and quantum Koszul resolution,''
  arXiv:hep-th/0602153.
  %%CITATION = HEP-TH 0602153;%%

\bibitem{Tan:2006qt}
  M.~C.~Tan,
  ``Two-dimensional twisted sigma models and the theory of chiral differential
  operators,''
  arXiv:hep-th/0604179.
  %%CITATION = HEP-TH 0604179;%%

\bibitem{Nekrasov:2005wg}
  N.~A.~Nekrasov,
  ``Lectures on curved beta-gamma systems, pure spinors, and anomalies,''
  arXiv:hep-th/0511008.
  %%CITATION = HEP-TH 0511008;%%
  
\bibitem{Witten:1987cg}
  E.~Witten,
  ``The index of the Dirac operator in loop space,''
  %
%\href{http://www.slac.stanford.edu/spires/find/hep/www?r=pupt-1050}{SPIRES entry}
{\it In Proc. of Conf. on Elliptic Curves and Modular Forms in Algebraic Topology, Princeton, N.J., (1986)}

\bibitem{Moore:1997dj}
  G.~W.~Moore, N.~Nekrasov and S.~Shatashvili,
  ``Integrating over Higgs branches,''
  Commun.\ Math.\ Phys.\ {\bf 209} (2000) 97--121,
  [arXiv:hep-th/9712241].
  %%CITATION = HEP-TH 9712241;%%

\bibitem{Nekrasov:2004vw}
  N.~Nekrasov and S.~Shadchin,
  ``ABCD of instantons,''
  Commun.\ Math.\ Phys.\  {\bf 252} (2004) 359
  [arXiv:hep-th/0404225].
  %%CITATION = HEP-TH 0404225;%%

\bibitem{Pressley:1988qk}
  A.~Pressley and G.~Segal,
  ``Loop Groups,'' Clarendon, Oxford, UK, (1988), 318 pp.
%\href{http://www.slac.stanford.edu/spires/find/hep/www?irn=1979159}{SPIRES entry}

\bibitem{Kreuzer:2006ax}
  M.~Kreuzer,
  ``Toric geometry and Calabi-Yau compactifications,''
  arXiv:hep-th/0612307.
  %%CITATION = HEP-TH 0612307;%%

\bibitem{Fulton:1993ab}
  W.~Fulton,
  ``Introduction to toric varieties,''
  Annals of Mathematics Studies, {\bf 131}, (1993) Princeton University Press 

\bibitem{Batyrev:1991ab}
  V.~Batyrev,
  ``On the classification of smooth projective toric varieties,''
  Tohoku Math. J. {\bf 43} (1991) 569--585. 

\bibitem{Szabo:1996md}
  R.~J.~Szabo,
  ``Equivariant localization of path integrals,''
  Lect.\ Notes\ Phys.\ {\bf M63} (2000) 1--319,
  [arXiv:hep-th/9608068].
  %%CITATION = HEP-TH 9608068;%%

\bibitem{Martelli:2006yb}
  D.~Martelli, J.~Sparks and S.~T.~Yau,
  ``Sasaki-Einstein manifolds and volume minimisation,''
  arXiv:hep-th/0603021.
  %%CITATION = HEP-TH 0603021;%%

\bibitem{Butti:2006au}
  A.~Butti, D.~Forcella and A.~Zaffaroni,
  ``Counting BPS baryonic operators in CFTs with Sasaki-Einstein duals,''
  arXiv:hep-th/0611229.
  %%CITATION = HEP-TH 0611229;%%

\bibitem{Forcella:2007wk}
  D.~Forcella, A.~Hanany and A.~Zaffaroni,
  ``Baryonic Generating Functions,''
  arXiv:hep-th/0701236.
  %%CITATION = HEP-TH 0701236;%%

\bibitem{Tan:2006by}
  M.~C.~Tan,
  ``The half-twisted orbifold sigma model and the chiral de Rham complex,''
  arXiv:hep-th/0607199.
  %%CITATION = HEP-TH 0607199;%%

\bibitem{Lust:2006zh}
  D.~Lust, S.~Reffert, E.~Scheidegger and S.~Stieberger,
  ``Resolved toroidal orbifolds and their orientifolds,''
  arXiv:hep-th/0609014.
  %%CITATION = HEP-TH 0609014;%%

\bibitem{Berkovits:2004px}
  N.~Berkovits,
  ``Multiloop amplitudes and vanishing theorems using the pure spinor
  formalism for the superstring,''
  Phys.\ Rev.\ D {\bf 66} (2002) 010001
  arXiv:hep-th/0406055.
  %%CITATION = HEP-TH 0406055;%%

\bibitem{Feigin:1990ut}
  B.~L.~Feigin and E.~V.~Frenkel,
  ``Representations Of Affine Kac-Moody Algebras And Bosonization,''
  in Physics and mathematics of strings, L.~Brink et al. editors, 271-316, (1990).

\bibitem{Malikov:1989ab}
  F.~G.~Malikov, 
  ``Singular vectors in Verma modules over affine algebras,'' 
  Funct.\ Anal.\ Appl.\ {\bf 23} (1989) 66--67

\bibitem{Berkovits:2002zk}
  N.~Berkovits,
  ``ICTP lectures on covariant quantization of the superstring,''
  Phys.\ Rev.\ D {\bf 66}, 010001 (2002)
  arXiv:hep-th/0209059.
  %%CITATION = HEP-TH 0209059;%%

\bibitem{Grassi:2005sb}
  P.~A.~Grassi and N.~Wyllard,
  ``Lower-dimensional pure-spinor superstrings,''
  JHEP {\bf 0512}, 007 (2005)
  [arXiv:hep-th/0509140].
  %%CITATION = HEP-TH 0509140;%%

\bibitem{Wyllard:2005fh}
  N.~Wyllard,
  ``Pure-spinor superstrings in d = 2, 4, 6,''
  JHEP {\bf 0511}, 009 (2005)
  [arXiv:hep-th/0509165].
  %%CITATION = HEP-TH 0509165;%%

\bibitem{Adam:2006bt}
  I.~Adam, P.~A.~Grassi, L.~Mazzucato, Y.~Oz and S.~Yankielowicz,
  ``Non-critical pure spinor superstrings,''
  arXiv:hep-th/0605118.
  %%CITATION = HEP-TH 0605118;%%

\bibitem{Krotov:2006th}
  D.~Krotov and A.~Losev,
  ``Quantum field theory as effective BV theory from Chern-Simons,''
  arXiv:hep-th/0603201.
  %%CITATION = HEP-TH 0603201;%%

\bibitem{Gotz:2006qp}
   G.~Gotz, T.~Quella and V.~Schomerus,
   ``The WZNW model on PSU$(1,1|2)$,''
   arXiv:hep-th/0610070.
   %%CITATION = HEP-TH 0610070;%%

\bibitem{Grassi:2004tv}
  P.~A.~Grassi and G.~Policastro,
  ``Super-Chern-Simons theory as superstring theory,''
  arXiv:hep-th/0412272.
  %%CITATION = HEP-TH 0412272;%%

\bibitem{Witten:1993yc}
  E.~Witten,
  ``Phases of N = 2 theories in two dimensions,''
  Nucl.\ Phys.\ B {\bf 403} (1993) 159
  [arXiv:hep-th/9301042].
  %%CITATION = HEP-TH 9301042;%%

\bibitem{Frenkel:2005ku}
  E.~Frenkel and A.~Losev,
  ``Mirror symmetry in two steps: A-I-B,''
  Commun.\ Math.\ Phys.\ {\bf 269} (2007) 39--86
  [arXiv:hep-th/0505131.]
  %%CITATION = HEP-TH 0505131;%%

\bibitem{Frenkel:2006fy}
  E.~Frenkel, A.~Losev and N.~Nekrasov,
  ``Instantons beyond topological theory. I,''
  arXiv:hep-th/0610149.
  %%CITATION = HEP-TH 0610149;%%

% \bibitem{Benvenuti:2006qr}
%   S.~Benvenuti, B.~Feng, A.~Hanany and Y.~H.~He,
%   ``Counting BPS operators in gauge theories: Quivers, syzygies and
%   plethystics,''
%   arXiv:hep-th/0608050.
%   %%CITATION = HEP-TH 0608050;%%

% \bibitem{Frenkel:2003np}
%   E.~Frenkel and M.~Szczesny,
%   ``Chiral de Rham Complex and Orbifolds,''
%   arXiv:math.ag/0307181.
%   %%CITATION = MATH-AG 0307181;%%

% \bibitem{Feng:2007ur}
%   B.~Feng, A.~Hanany and Y.~H.~He,
%   ``Counting gauge invariants: The plethystic program,''
%   arXiv:hep-th/0701063.
%   %%CITATION = HEP-TH 0701063;%%

   

\end{thebibliography}
\end{document}